\documentclass[preprint,pteplogo]{ptephy_v2}

\preprintnumber{XXXX-XXXX} 
\usepackage{hyperref}

\usepackage{amsmath} 

\begin{document}

\title{Polarization-dependent observables in $H\to \ell^{+}\ell^{-} \gamma$ in the SM}


\author[1]{Usman Hasan}
\affil{National Centre for Physics, Shahdra Valley Road, Islamabad 44000, Pakistan. \email{usman.hasan@ncp.edu.pk, ambreen.uzair@ncp.edu.pk, mjunaid@ncp.edu.pk, bilal.tariq@ncp.edu.pk, ishtiaq.ahmed@ncp.edu.pk, shahin.iqbal@ncp.edu.pk}}
\author[1]{A. Uzair}
\author[1]{M. Junaid}
\author[1]{Bilal Tariq} 
\author[1]{Ishtiaq Ahmed}
\author[1]{Shahin Iqbal}


\begin{abstract}%
The rare three body decay of a Higgs boson  to a lepton-anti lepton pair and a photon has begun to attract attention, after the first evidence for the $H\to Z\gamma$ at CMS and ATLAS, which is a sub process of $H \to \ell^+ \ell^- \gamma$ . To investigate some important features of this process, we suggest that the polarized forward-backward and the photon polarization asymmetries could be useful to probe its important properties, such as the behavior of Yukawa coupling, resonance, and non-resonance contributions. Our analysis introduces a comprehensive framework to evaluate the aforementioned polarization-dependent observables. By analyzing the polarization effects of the final-state photon and lepton separately on forward-backward asymmetries, we demonstrate that loop-induced contributions play a significant role to investigate these asymmetries. Unlike the unpolarized case, where the interference effects of resonance and non-resonance effects are minimal, we show that polarization dependent observables offer a powerful tool to analyze these features of this decay mode. Furthermore, these observables can provide a handy tool for probing possible signatures of physics beyond the SM.
\end{abstract}

\subjectindex{xxxx, xxx}

\maketitle

\section{Introduction}
The 2012 discovery of the Higgs particle by the ATLAS \cite{ATLAS:2012yve} and CMS \cite{CMS:2012qbp} experiments marked a significant milestone in confirming the Glashow-Salam-Weinberg (GSW) theory of weak interactions. GSW, which forms the $SU(2)\otimes U(1)$ part of the Standard Model (SM), unifies the weak and electromagnetic interactions. Now, the next goal in current and future colliders, such as the High Luminosity Large Hadron Collider (HL-LHC) and the International Linear Collider (ILC) \cite{Liss:2013hbb, CMS:2013xfa}, is to precisely measure properties of the Higgs particle discovered by the LHC \cite{ATLAS:2012yve,CMS:2012qbp} and determine if it is indeed the SM Higgs boson, or otherwise .Quantities like branching ratios, forward-backward asymmetries, and polarization asymmetries can be useful probes for answering this question. In this context, the Higgs couplings to $W$, $Z$, $\tau$, $b$, $t$, and photons have been measured by the CMS \cite{CMS:2017zyp, CMS:2018zzl, CMS:2019ekd, CMS:2013fjq, CMS:2018nsn, CMS:2018fdh, CMS:2018piu} and ATLAS \cite{ATLAS:2015muc, ATLAS:2014aga, ATLAS:2014xzb, ATLAS:2018pgp, ATLAS:2015xst, ATLAS:2018ynr, ATLAS:2018kot, ATLAS:2018mme, ATLAS:2018hxb} collaborations. However, current data, such as the CMS results at $150 \text{ fb}^{-1}$, are insufficient to measure spin asymmetries \cite{CMS:2022ahq, CMS:2023mku}. The technical challenges in measuring these asymmetries are expected to diminish with advancements in experimental capabilities. The HL-LHC, with its projected luminosity increase to $450-3000 \text{ fb}^{-1}$, along with future $e^{+}e^{-}$ colliders, should enable better asymmetry measurements \cite{Schmidt:2016jra, Rossi:2019swj}.

The semi-leptonic Higgs decay has gained a lot of attention due to experimental accessibility at the Large Hadron Collider (LHC) and the upcoming high-luminosity LHC \cite{CMS:2019ajt, ATLAS:2019cid, ATLAS:2018bnv, Fonseca:2016tbn, Curtin:2013fra, Sun:2013cba,CMS:2022ahq,Sun:2013cba, Hernandez-Juarez:2024pty, Kachanovich:2024vpt}. Both CMS and ATLAS have recently presented first evidence of $H\to Z\gamma$ channel through its leptonic decays $(Z\to\ell^+\ell^-,\text{ } \ell=e \text{ or }\mu)$ which is a sub-process of $H\rightarrow\ell^+ \ell^-\gamma$ decays as $H\to Z^*[\to\ell^+\ell^-]\gamma$ \cite{CMS:2023mku}. Therefore, this channel can work as a tool to test the SM properties of the Higgs boson at high energy, which are recently becoming accessible at current and future experimental facilities. In addition, this channel may also provide an additional avenue for indirect searches to probe any possible new physics (NP) i.e., the physics beyond the SM \cite{Sun:2013cba,Krause:2018wmo,Athron:2021kve,Tran:2023vgk,Benbrik:2022bol,Denner:2019fcr,Kanemura:2017gbi,Kanemura:2019slf,Kanemura:2022ldq,Aiko:2023xui,Phan:2021xwc,VanOn:2021myp,Kachanovich:2020xyg,Hue:2023tdz,Chiang:2012qz} as many new particles can propagate in the loop diagrams of these decay processes. Similarly, the precise measurements of these decays may also explore any mixing of new neutral Higgs bosons with the SM-like Higgs bosons.

From the theoretical point of view, the $H\rightarrow\ell^+\ell^-\gamma$ decay  receives contributions from tree and loop levels diagrams. However, for $\ell=e$, the tree level contribution is much smaller than the loop level contribution while for $\ell=\mu$ case, the tree and loop contributions are comparable. The loop contributions in this process are further classified in three types of diagrams: (i) the diagrams with off-shell boson, (ii) the diagrams with off-shell photon and (iii) the box diagrams. Regarding this, calculations up to one loop level have been carried out by several groups \cite{Passarino:2013nka,Kachanovich:2020xyg,Abbasabadi:1996ze,Dicus:2013ycd,Han:2017yhy} and recently calculated in $R_\xi$ \cite{Kachanovich:2020xyg} and having the comparison with previous studies in ref. \cite{Kachanovich:2021pvx}.

In addition, it was mentioned in \cite{Kachanovich:2021pvx}, that  there are discrepancies in the calculations of the decay rates of $H\to\ell^+\ell^-\gamma$  due to the choice of QED fine structure constant $\alpha$. The forward-backward asymmetry, $A_{FB}$, is also calculated in the centre of  mass frame in \cite{Korchin:2014kha} and \cite{Sun:2013cba} where the results are at variance. Ref. \cite{Sun:2013cba} finds  $A_{FB}$ to be zero, while ref. \cite{Korchin:2014kha} finds it be to non-zero.  Kachanovich et. al.  reported nonzero $A_{FB}$ in the rest frame of Higgs boson \cite{Kachanovich:2021pvx}. Similarly, the polarization asymmetries are calculated in \cite{Akbar:2014pta}, where the box diagrams were missing which were addressed recently in \cite{shahkar}.
In the current study, we have calculated some other interesting observables such as the decay rate of $H\to\ell^+\ell^-\gamma$ when photon is polarized, $A_{FB}$ in the semi-leptonic Higgs decay $H\to \ell^{+}\ell^{-}\gamma$ with various different final state lepton and photon polarization configurations. We found that these  asymmetries also have important features which provide complimentary tool to explore the further properties of this channel.

The motivation to calculate the above mentioned observables is twofold: Firstly, as discussed above,  the previous studies  \cite{Passarino:2013nka,Kachanovich:2020xyg,Abbasabadi:1996ze,Dicus:2013ycd,Han:2017yhy}  show that the calculation of the decay rates of $H\to\ell^+\ell^-\gamma$  depends on the QED fine structure constant, $\alpha$. As the asymmetries are, in general, ratios of the decay rates, particularly in the case of the electron, the factors of $\alpha$ almost completely cancel out.  Therefore, the asymmetries may be regarded as good complementary observables in addition to differential decay rates to further explore the properties of this channel.  Secondly, as reported in the previous study \cite{shahkar} where an anomalous behavior was reported around 60 GeV, in the decay rate and the polarization asymmetry of the lepton when the final-state lepton is longitudinally polarized. This behavior is explained by the inclusion of the effect of Z-resonance contribution \cite{Kachanovich:2021pvx} which provides an opportunity to explore the properties of the $H\to Z\gamma$ sub process in  $H\to\ell^+\ell^-\gamma$ decays. It is therefore of particular interest to see whether a similar anomalous behaviour is present in other asymmetries such as $A_{FB}$ and photon polarization asymmetries, $A_P$, which are analyzed in the current study.

This paper is organized as follows: In section \ref{theory} , we formulate the theoretical framework for the process $H\rightarrow \ell^{+}\ell^{-}\gamma$, required for the calculations and present the formulae of the $A_{FB}$ and $A_P$. In section \ref{analysis}, we describe our results and  the phenomenological analysis of the observables studied in the present work. Finally in section \ref{con}, we present a brief summary and conclusion of our work.

\verb+ptephy_v1.cls+ v0.1

\section{Theoretical Framework}\label{theory}

In this section, we describe the theoretical framework for calculating the observables under consideration. The tree and the loop level amplitude for the process, $H\to\ell^+\ell^-\gamma$, where $\ell= e,\mu$, can be expressed as follows \cite{Sun:2013rqa,Kachanovich:2020xyg,Kachanovich:2021pvx,VanOn:2021myp},
\begin{eqnarray}
\mathcal{M}_{tree}&=&\mathcal{C}_{0}
\bar{u}(p_2)\bigg(\frac{2p_{2}^{\nu}+\gamma^{\nu}\not{k}}{2p_{2}\cdot k}-\frac{\not k \gamma^{\nu}+2p_{1}^{\nu}}{2p_{1} \cdot k}
\bigg)v(p_1)\epsilon_{\nu}^{*}\, \label{MT}\\
\mathcal{M}_{loop}&=&\epsilon^{*\nu}\{(k_{\mu}p_{1\nu}-g_{\mu\nu}(k.p_1))\bar{u}(p_2)(\mathcal{C}_{1}\gamma^{\mu}+\mathcal{C}_{2}\gamma^{\mu}\gamma^{5})v(p_{1})\notag\\
&&+(k_{\mu}p_{2\nu}-g_{\mu\nu}(k.p_2))\bar{u}(p_2)(\mathcal{C}_{3}\gamma^{\mu}+\mathcal{C}_{4}\gamma^{\mu}\gamma^{5})v(p_1)\}.\label{ML}
\end{eqnarray}
Here $p_1,p_2$ and $k$ are the 4-momenta of the final state lepton, anti-lepton and the photon, while $\epsilon^{\nu}$ is the photon poloarization vector. 
 $2p_2\cdot k=u-m_{\ell}^2$ and $2p_1\cdot k=t-m_{\ell}^2$ where $s,t,u$ are the Mandelstam variables defined as $ m_{\ell\ell}^2\equiv s=(p_1+p_2)^2$, $ m_{\ell\gamma}^2\equiv t=(p_1+k)^2$ and $u=(p_2+k)^2$ and coefficient $\mathcal{C}_{0}=-\frac{4\pi \alpha
 m_\ell}{m_{W} sin\theta_{W}}$. The coefficients $\mathcal C_{1},\mathcal C_{2},\mathcal C_{3}$ and $\mathcal C_{4}$ can be expressed as,
\begin{eqnarray}
\mathcal C_{1}&=&\frac{a_1+b_1}{2} \qquad \qquad
\mathcal C_{2}=\frac{a_1-b_1}{2},\notag \\
\mathcal C_{3}&=&\frac{a_2+b_2}{2} \qquad \qquad \mathcal C_{4}=\frac{a_2-b_2}{2}.\label{Cfunctions}
\end{eqnarray}
Here $a_i,b_i$ (where $i=1,2$) are functions of Mandelstam variables $s, t, u$ and are given in terms of the Passarino-Veltman decomposition of the tensor integrals \cite{Kachanovich:2021pvx} calculated in the limit of mass-less leptons. 
\begin{figure*}[]
\centering\scalebox{1}{
\begin{tabular}{cc}
\includegraphics[width=3in,height=2in]{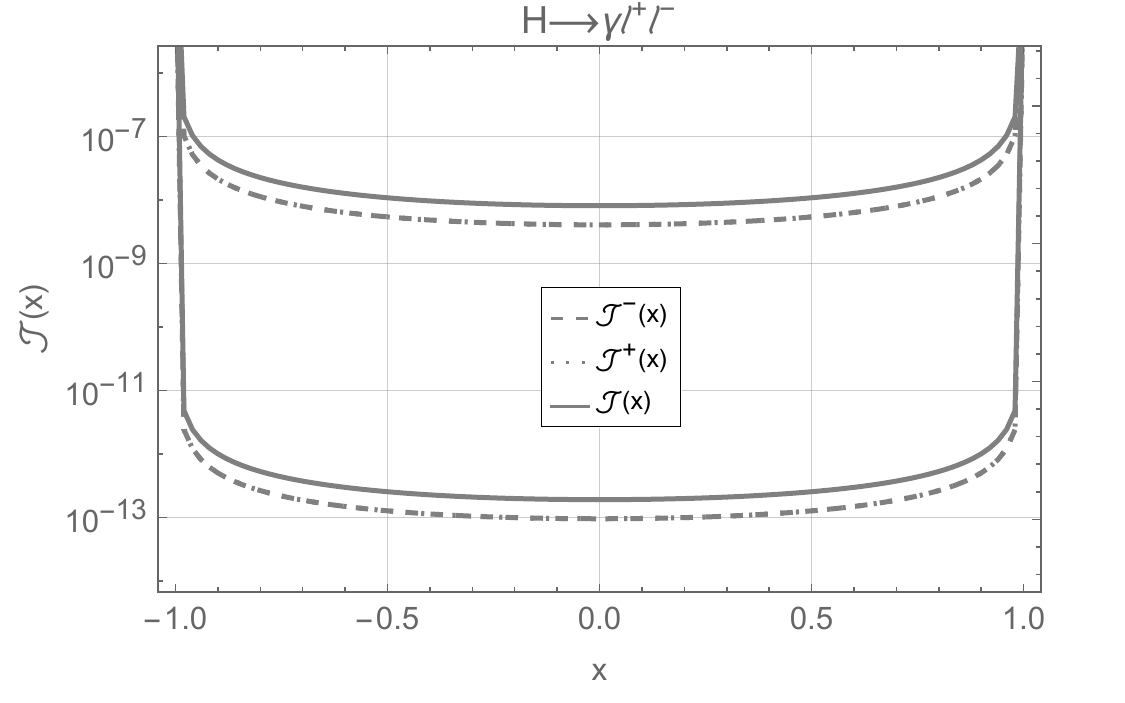}&
\includegraphics[width=3in,height=2in]{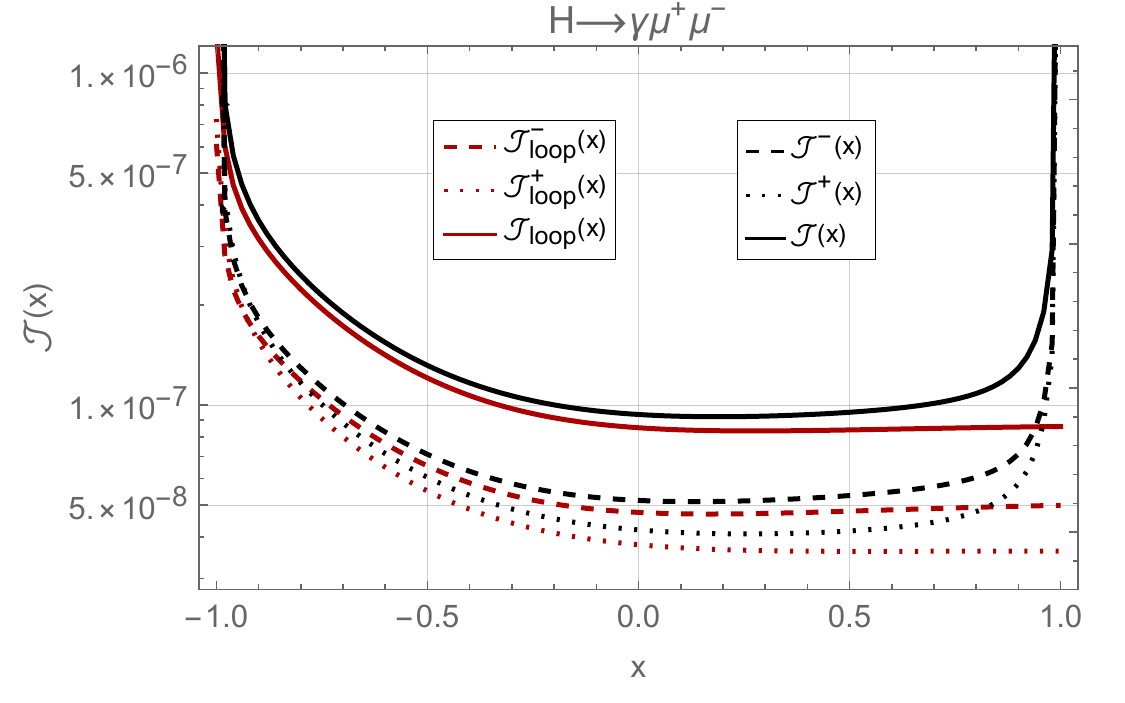} \\
(a) & (b) \\
\includegraphics[width=3in,height=2in]{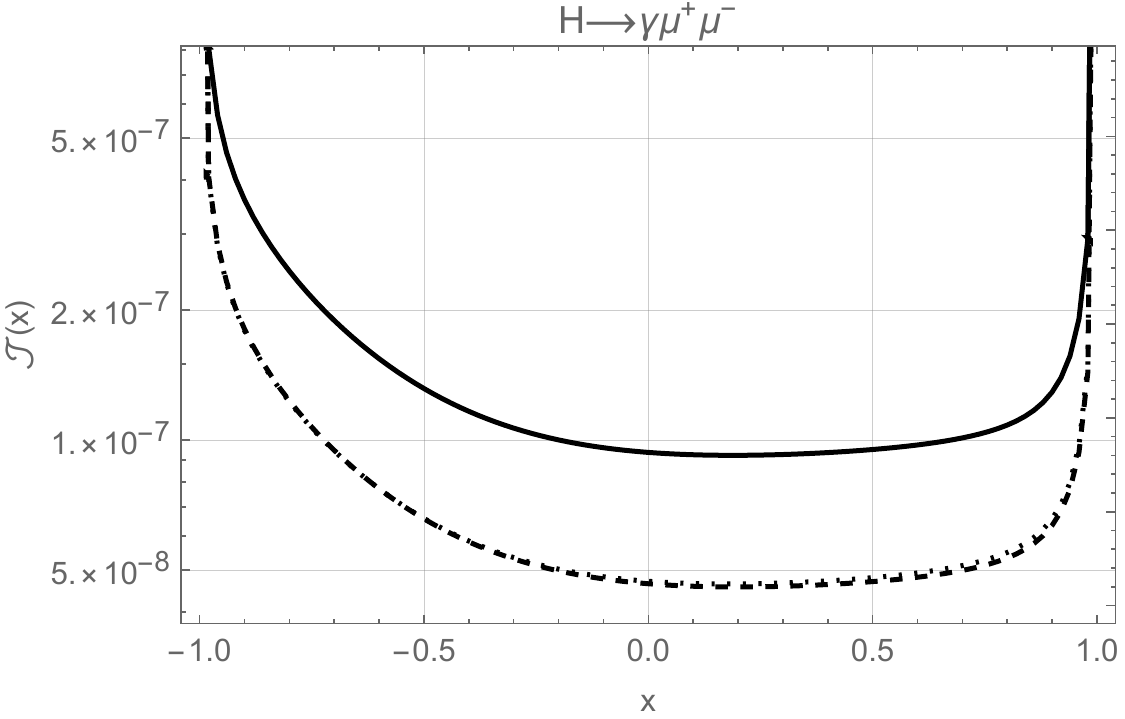}&
\includegraphics[width=3in,height=2in]{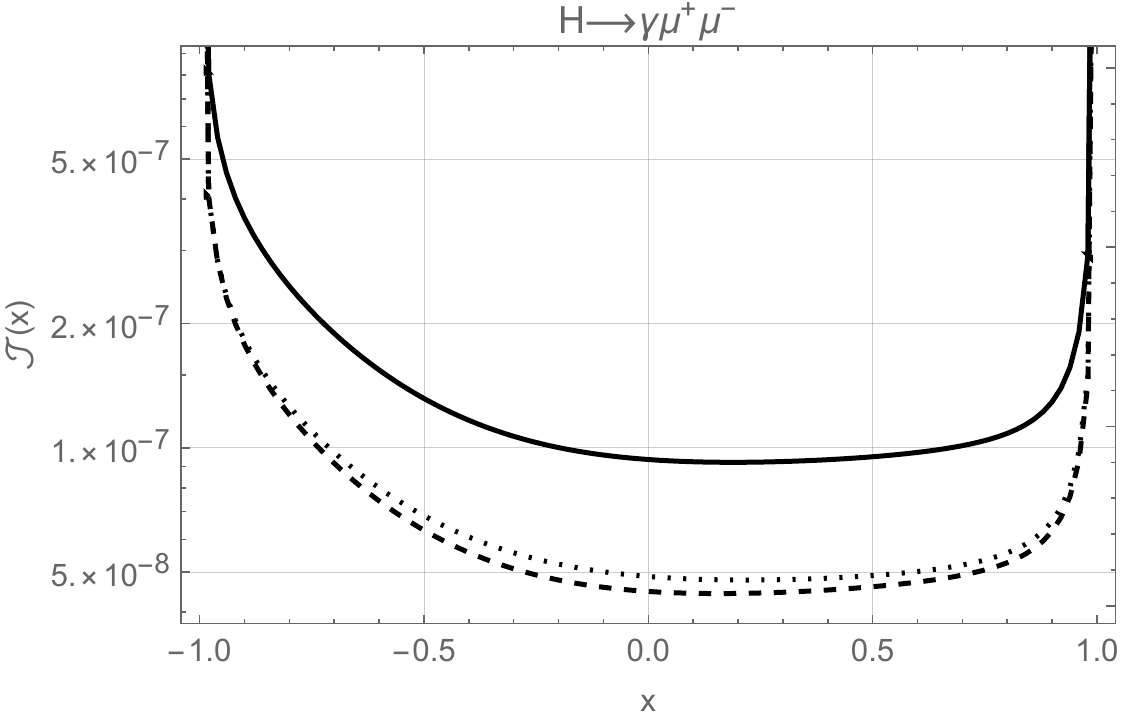}\\
(c)&(d) \\
\includegraphics[width=3in,height=2in]{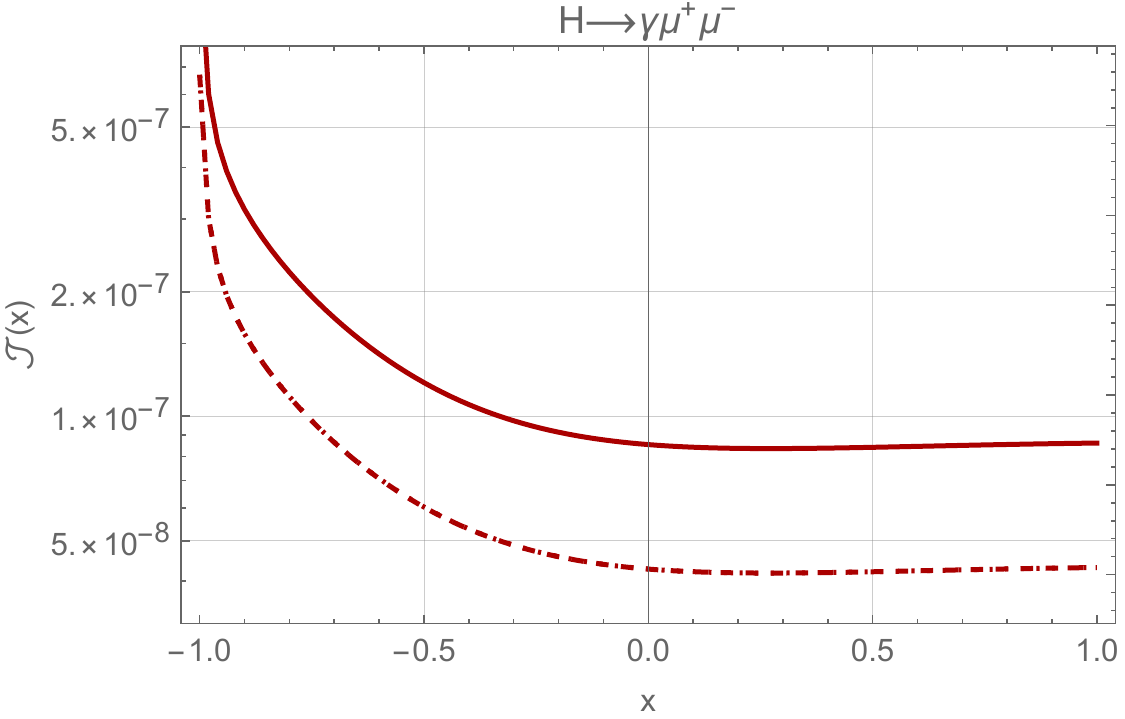}&
\includegraphics[width=3in,height=2in]{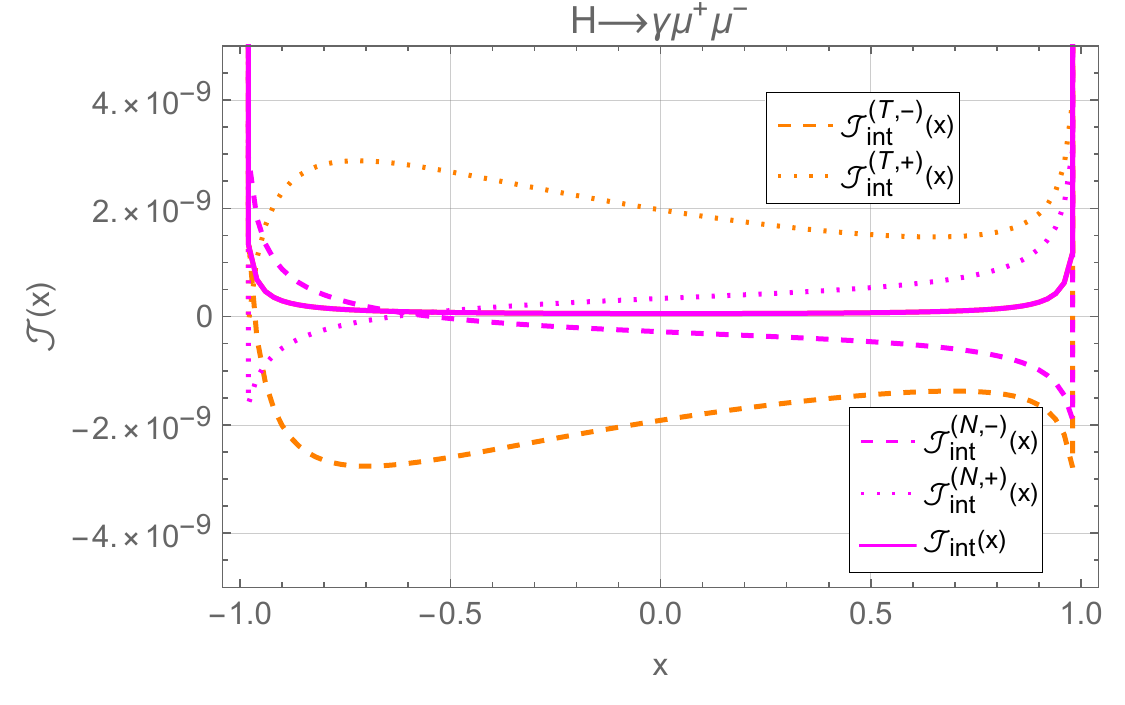}\\
(e) & (f)
\end{tabular}}
\caption{Lepton-polarized and unpolarized Decay Rates $\mathcal{J}^{(i,\pm)}(x)$ against $x$. Solid lines show unpolarized while dotted (dashed) lines show positively (negatively) polarized decay rates. The black, red and grey curves show the total, the loop and the tree contributions respectively.
(a) Tree contributions where the upper curve is for muon and the lower for electron. (b) Total and loop contributions for longitudinally polarized muon. (c) Total contribution of normally polarized muon. (d) Total contribution of transversely polarized muon. (e) Loop contribution of normally and transversely polarized muon. (f) Interference term of tree and loop contributions for the normally (purple curves) and transversely (orange curves) polarized muon.}
\label{LPDRx}
\end{figure*}

\subsection{Polarized Decay Rates}
In the current study, we have analyzed the decay rates of the $H\to\ell^+\ell^-\gamma$ process in the Higgs rest frame when the final-state lepton (photon) is polarized by using the polarization vectors given in Appendix A:
 We will write the unpolarized decay rate as
 \begin{eqnarray}
 \mathcal{J}(s,x)\equiv\frac{d^2\Gamma(s,x)}{dxds},
 \end{eqnarray}
the photon-polarized decay rate as
 \begin{eqnarray}
 \mathcal{J}^{(\pm)}(s,x)\equiv\frac{d^2\Gamma^{(\pm)}(s,x)}{dxds},
 \end{eqnarray}
and the lepton-polarized decay rate as
 \begin{eqnarray}
 \mathcal{J}^{(i,\pm)}(s,x)\equiv\frac{d^2\Gamma^{(i,\pm)}(s,x)}{dxds}.\notag\\
 \end{eqnarray}
   Here $ x\equiv\cos{\theta}$, $i=L, N,$ and $T$ represent longitudinal, normal, and transverse polarization of the final state leptons, respectively, and $\pm$ correspond to their spins  $\pm\frac{1}{2}$.

As $ t=m_{\ell}^2+2 E_\gamma \left(E_{\ell}-x p_{\ell} \right)$, therefore, $\mathcal{J}^{(i,\pm)}(s,x)$ can be found by 


\begin{eqnarray}
 \frac{d^2\Gamma^{(i,\pm)}(s,t)}{dsdt} =\frac{1}{256\pi^3m_{H}^3}\vert\mathcal{M}^{(i,\pm)}(s,t)\vert^2.\label{2.10}
\end{eqnarray}
 Here $+(-)$ corresponds to the decay rate when final state lepton has $+\frac{1}{2}(-\frac{1}{2})$ polarization and $\mathcal{M}^{(i,\pm)}(s,t)=\mathcal{M}^{(i,\pm)}_{tree}(s,t)+\mathcal{M}^{(i,\pm)}_{loop}(s,t)$. The expressions of $\vert\mathcal{M}^{(i,\pm)}(s,t)\vert^2$ are given in \cite{shahkar}. Here
 \begin{eqnarray}
         p_{\ell}&=&\sqrt{E_{\ell}^2-m_{\ell}^2}, \qquad E_{\ell}=\frac{s+t-m_{\ell}^2}{2 m_H}.
 \end{eqnarray}
 where $p_{\ell},E_{\ell},m_{\ell}$ are respectively the momentum, energy, and mass of the final state lepton.
 After integrating on $s\text{ (or }x)$, one can find the decay rate $\mathcal{J}$ as a function of $s \text{ or } x \text{ } (s/x )$
 i.e. $\mathcal{J}(s/x)\equiv\int\mathcal{J}(s,x)ds/dx$.
 To get the numerical results of observables, we set the following limits for the Mandelstam variables $s$ and $t$:
\begin{eqnarray}
   &&s_{min}=4m_\ell^2,\qquad\qquad s_{max}=m_H^2,\notag \\
    &&t_{_{max}^{min}}=\frac{1}{2}\left(m_H^2-s+2m_\ell^2\mp(m_H^2-s)\sqrt{1-4m_\ell^2/s}\right),\notag \\\
    \end{eqnarray}
 where the numerical values of various parameters are given in Table \ref{Numericals}. We have retained the finite masses of leptons in the phase space integration. In Table \ref{wc table}, we have calculated the numerical values of polarized decay rates by using the following kinematic cuts \cite{Kachanovich:2021pvx,Kachanovich:2020xyg}:
 \begin{eqnarray}
    E_{\gamma , min}&=&5\text{ Gev},\qquad\qquad s,t,u>(0.1m_H)^2.
    \end{eqnarray}

\subsection{Forward-Backward and Photon Polarization Asymmetries}
In addition to the polarized decay rate, we have analyzed the unpolarized $A_{FB}$ and lepton-polarized (photon-polarized) forward-backward asymmetries $A_{FB}^{i,\pm}$ ($A_{FB}^\pm$). Furthermore, we have also studied the photon polarization asymmetries, $A_P(x)$ and $A_P(m_{\ell\gamma})$.
The formulas for these observables are defined as follows:

 \begin{eqnarray}
 A_{FB}^{(i,\pm)}(s)=\frac{\int_{-1}^0\mathcal{J}^{(i,\pm)}(s,x)dx-\int_{0}^{1}\mathcal{J}^{(i,\pm)}(s,x)dx}{\int_{-1}^{1}\mathcal{J}(s,x)dx}.\notag \\\label{AFBformula}
 \end{eqnarray} 
 
Forward-backward asymmetry when the final state photon is polarized \cite{Korchin:2014kha,Colangelo:2023xnu}:
\begin{eqnarray}
 A_{FB}^{(\pm)}(s)=\frac{\int_{-1}^0\mathcal{J}^{(\pm)}(s,x)dx-\int_{0}^{1}\mathcal{J}^{(\pm)}(s,x)dx}{\int_{-1}^{1}\mathcal{J}(s,x)dx}.
 \end{eqnarray}

Lastly, photon polarization asymmetry is defined as \cite{Gabrielli:2005ek,Colangelo:2023xnu}:
\begin{eqnarray}
A_p(x) =\frac{\int\mathcal{J}^{(+)}(s,x)ds-\int\mathcal{J}^{(-)}(s,x)ds}{\int\mathcal{J}(s,x)ds}.
\label{sin}
\end{eqnarray}

\begin{table*}[htb]
\begin{tabular}{ccc}
\hline\hline
$\alpha^{-1}=132.184$, $m_{H}=125.1$ GeV, & $m_W=80.379$ GeV, & $m_Z=91.1876$ GeV,\\
$m_{t}=173.1$ GeV, & $m_{\mu}=0.106$ GeV, & $m_e=0.51\times10^{-3}$ GeV,\\ $\Gamma_Z=2.4952$ GeV,  & $C_W=0.881469$, & $G_F=1.1663787\times10^{-5}$ GeV$^{2}$.\\
\hline\hline
\end{tabular}\label{input}
\centering \caption{Default values of input parameters used in the
calculations.}
\label{Numericals}
\end{table*}

\section{Phenomenological Analysis}\label{analysis}
In this study, we have calculated (i) the decay rates of $H\to\ell^+\ell^-\gamma$ against $x$ when the final state lepton or photon is polarized, (ii) the decay rates against $m_{\ell\gamma}$, $\mathcal{J}^{(\pm)}(t)$, when only the photon is polarized, (iii) the photon polarization asymmetry $A_{p}(x)$, (iv) the  lepton (photon) polarized forward-backward asymmetry $A_{FB}^{(i,\pm)}(s) (A_{FB}^{(\pm)}(s))$. The phenomenological analysis of these observables is described separately in the following sections.

\subsection{Lepton-Polarized Decay Rates $\mathcal{J}^{(i,\pm)}$}
In Fig. \ref{LPDRx}, the decay rates $\mathcal{J}^{(i,\pm)}(x)$ of $H\rightarrow\ell^{+}\ell^{-}\gamma$ are shown when the final-state lepton is longitudinally, transversely, or normally polarized.  Before the analysis, we first describe the color scheme for our plots, unless defined otherwise. The solid lines show unpolarized decay rates, while the dotted (dashed) lines show positively (negatively) polarized decay rates. The black, red and gray curves show the total, loop, and tree contributions of decay rates, respectively.

Fig. \ref{LPDRx} (a) shows the tree contribution of the decay rates, where the upper curve is for the muon and the lower curve for the electron. It is observed that no effect of polarization is found at the tree level whether the photon or the lepton is polarized. Furthermore, the muon contribution is greater than that of the electron because of a difference in the strength of the Yukawa coupling. The total and loop contributions for the longitudinally polarized muon are shown in Fig. \ref{LPDRx} (b), where we observe that the loop contributions are approximately equal for the case of electron and muon (red curves). It is also noted that the polarization effects appear at the loop level as the positively polarized decay rate $\mathcal{J}^{(L,+)}(x)$ is suppressed compared to the negatively polarized $\mathcal{J}^{(L,-)}(x)$. This difference increases as the angle between photon and lepton changes from $\theta=\pi$ to $\theta=0$. However, near $\theta=\pi$, the loop-level polarization effects disappear. While, near $\theta=0$, the tree-level contribution becomes dominant; therefore, the polarization effects are also suppressed in this region, as can be seen from Fig. \ref{LPDRx} (b). This inference can be used to confirm the Yukawa coupling strength in different kinematical regions.
In contrast, the normal polarization effects are negligible (see Fig. \ref{LPDRx} (c)), while as compared to longitudinal case, the transverse polarization effects on the decay rates are mild (see Fig. \ref{LPDRx} (d)). 

To explain further the origin of normal and transverse polarization effects on the decay rates, we have drawn the loop-level contributions in Fig. \ref{LPDRx} (e), where one can see that there is no polarization effect in the decay rates. Therefore, it is found that this polarization effect is not coming from the loop but originates from the interference of the tree and the loop contributions which is shown in Fig. \ref{LPDRx} (f). This behavior can be a complementary check to analyze the Yukawa coupling throughout the kinematic region. 

\subsection{Photon-Polarized Decay Rates $\mathcal{J}^{\pm}$}
\begin{figure*}[!htb]
\centering
\begin{tabular}{cc}
\includegraphics[width=3in,height=2in]{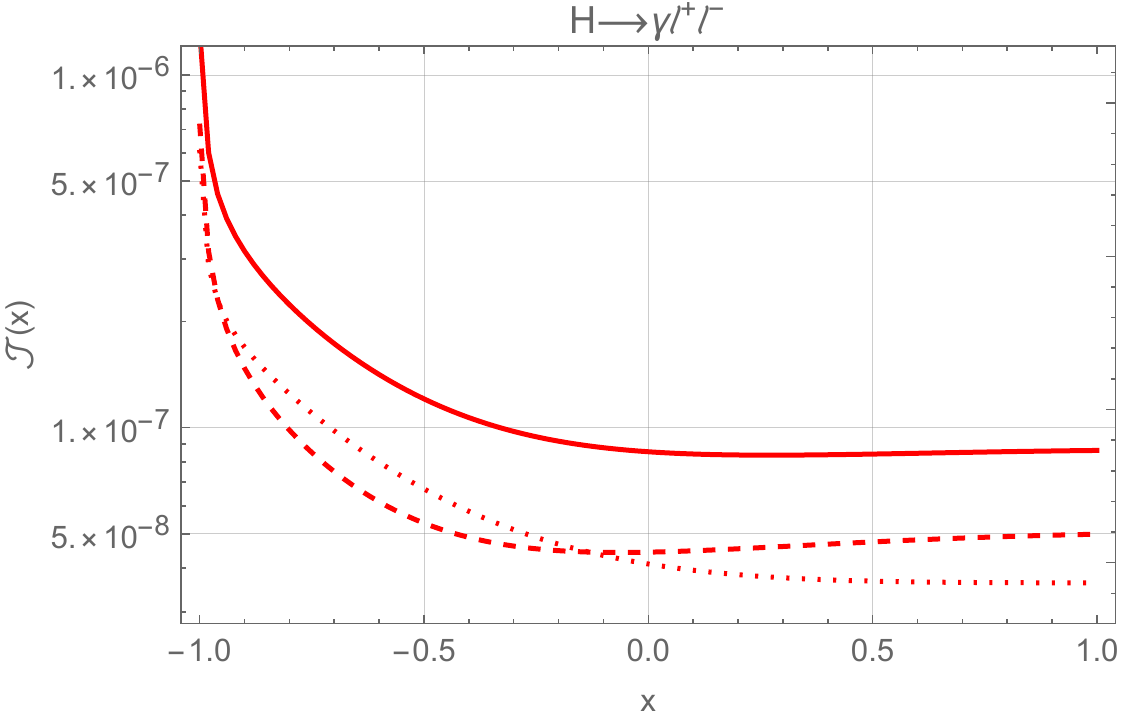}&
\includegraphics[width=3in,height=2in]{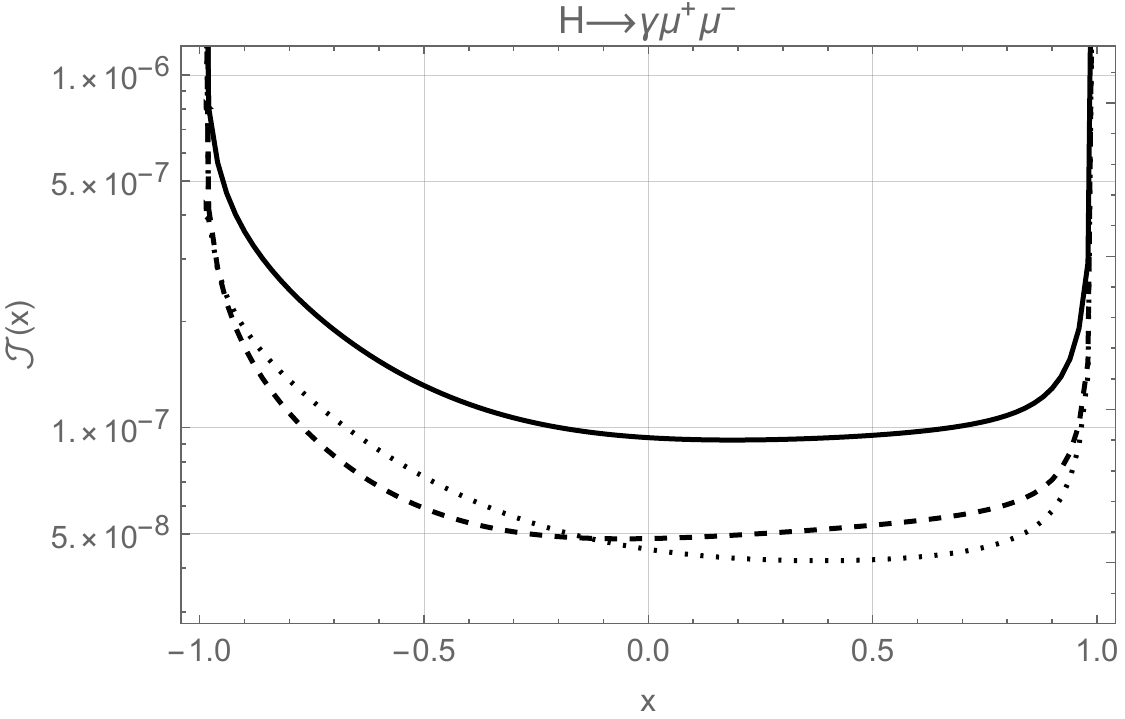}\\
(a) & (b) \\
\includegraphics[width=3in,height=2in]{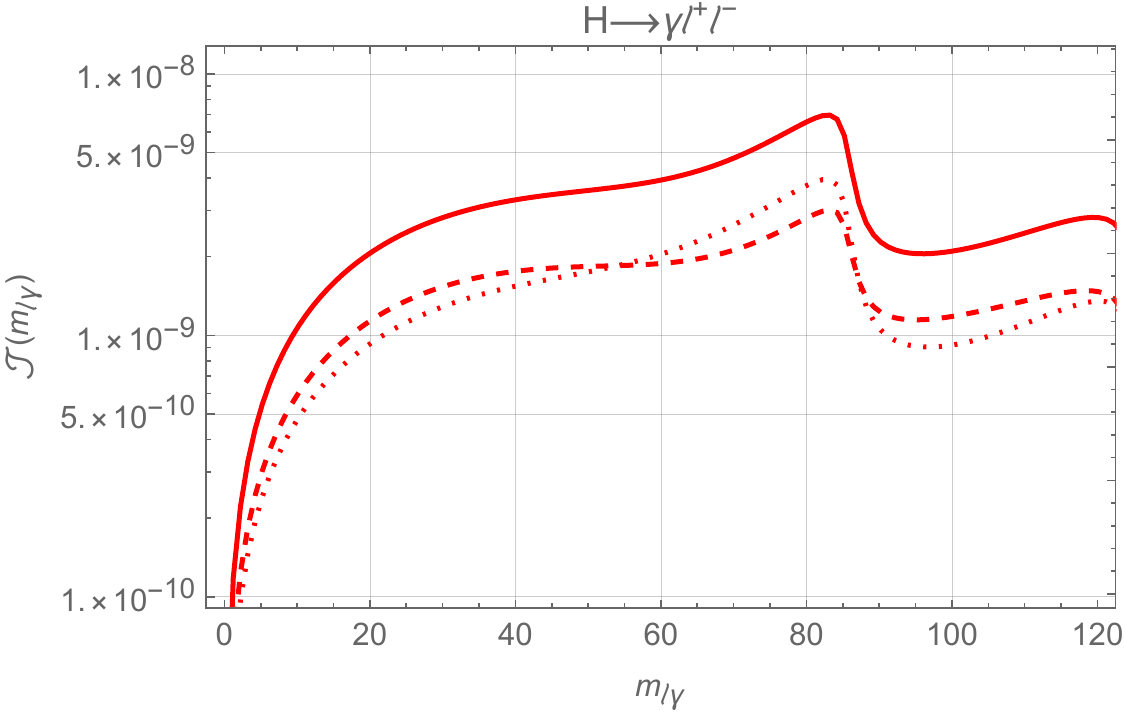}&
\includegraphics[width=3in,height=2in]{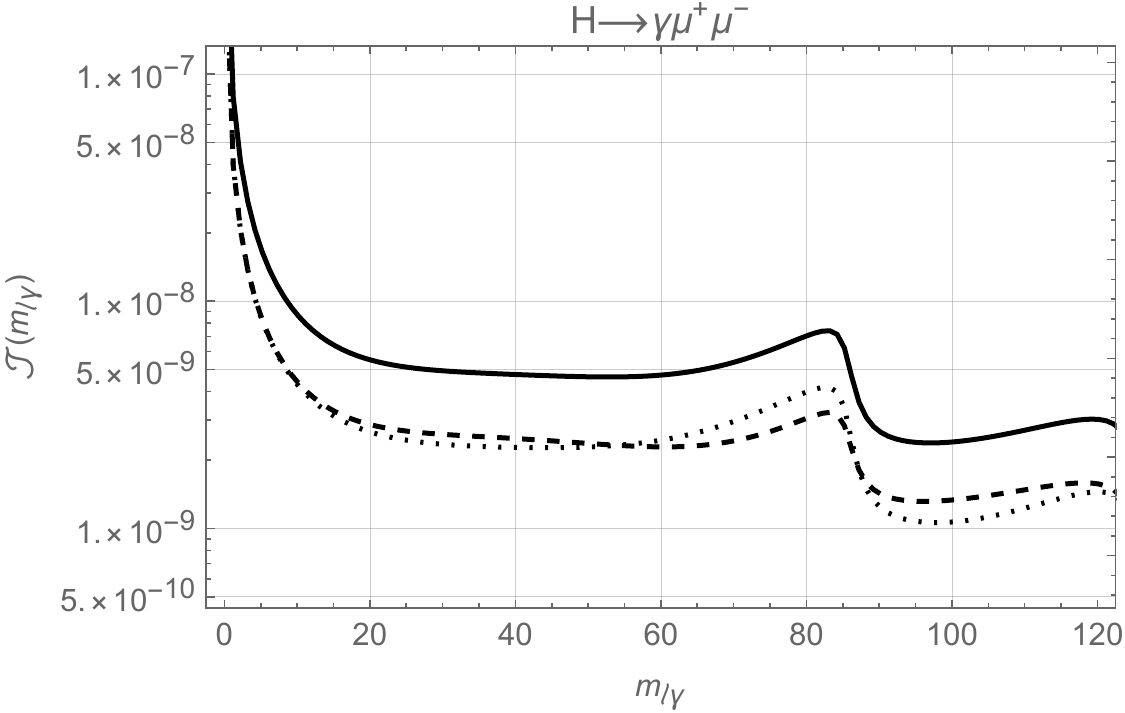}\\
(c) & (d) \\
\end{tabular}
\caption{Gamma polarized decay rates against $x$ and $m_{\ell\gamma}$ i.e., $\mathcal{J}^{(\pm)}(x)\text{ and }\mathcal{J}^{(\pm)}(m_{\ell\gamma})$. Here (a) and (c) show loop contributions and are valid for both the final state electron and muon cases whereas (b) and (d) represent total contributions only for muon. While the color scheme for solid, dashed and dotted curves is same as described in Fig. \ref{LPDRx}.}
\label{GPDR}
\end{figure*}

As mentioned above, there are no polarization effects at tree level, therefore, the Fig. \ref{LPDRx}(a), is  also valid for photon-polarized decay rates. It is worth mentioning here that the decay rates as a function of $m_{\ell\ell}$ are independent of photon polarization. However, the photon polarization effects to the decay rate are prominent at the loop level as shown in Fig. \ref{GPDR} which is similar to the case when the lepton is longitudinally polarized. In Fig. \ref{GPDR} (a), we observe that when lepton and photon are anti-parallel ($\theta\simeq\pi$), the polarization effects are negligible; however, for the final state lepton and photon are parallel  ($\theta\simeq0$), the polarization effects are prominent in the decay rates. In addition, at $\theta\simeq0.54 \pi$, the photon polarization effects disappear again after this point. Before this point, the positively polarized photon contribution (dotted curve) to the decay rate is higher than the negatively polarized photon contribution (dashed curve), whereas after this point, the behavior is reversed.

Fig. \ref{GPDR} (b) shows the total decay rates for muon where one can see near $\theta\simeq0$, similar to the lepton-polarized case, the tree-level contribution is dominant; therefore, the polarization effects become negligible.
To further analyze the effects of photon polarization, we have also plotted the decay rates against $m_{\ell\gamma}$ in Figs. \ref{GPDR} (c) and (d). Here, one can see that the polarization effects vanish near $m_{\ell\gamma}\simeq0, 60\text{ and } 120$ GeV. However, before $m_{\ell\gamma}\simeq60$ GeV, the negatively polarized photon contribution (dashed curve) to the decay rate is higher than the positively polarized photon contribution (dotted curve) and after this value, until resonance, the behavior is reversed. However, after resonance, the contribution to the decay rate of positive-photon polarization becomes dominant again.

\subsection{Photon Polarization Asymmetry $A_P$}
\begin{figure*}[htbp]
\centering
\begin{tabular}{cc}
\includegraphics[width=3in,height=2in]{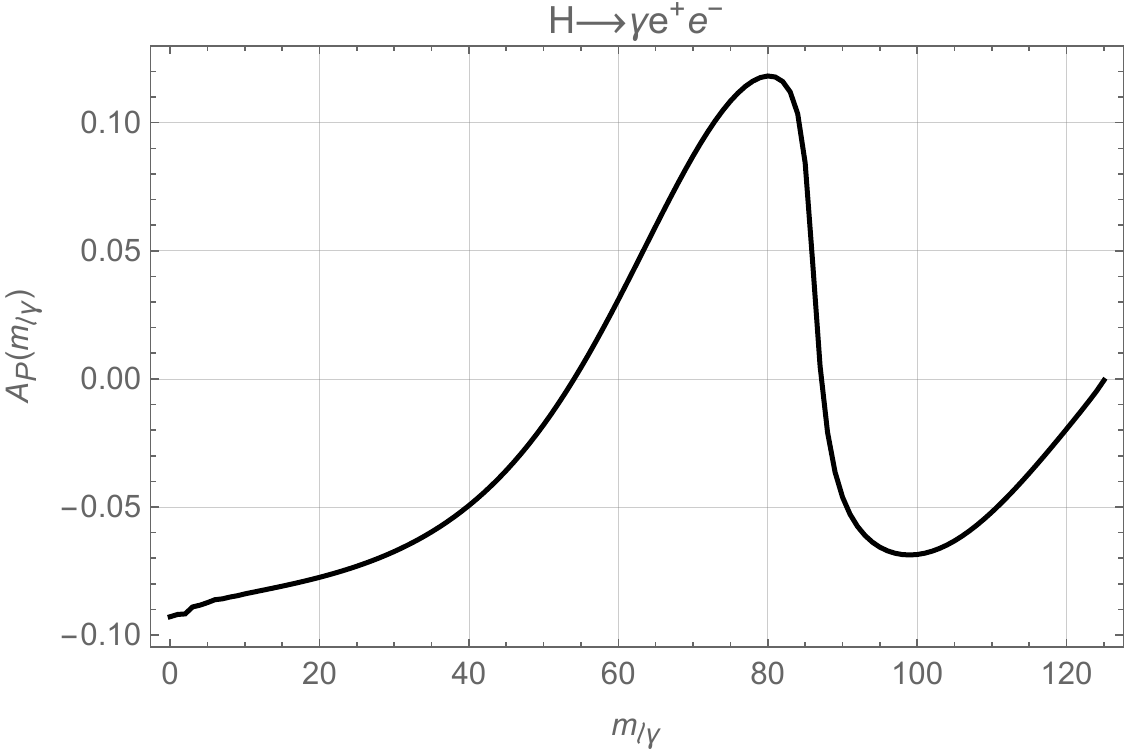}&
\includegraphics[width=3in,height=2in]{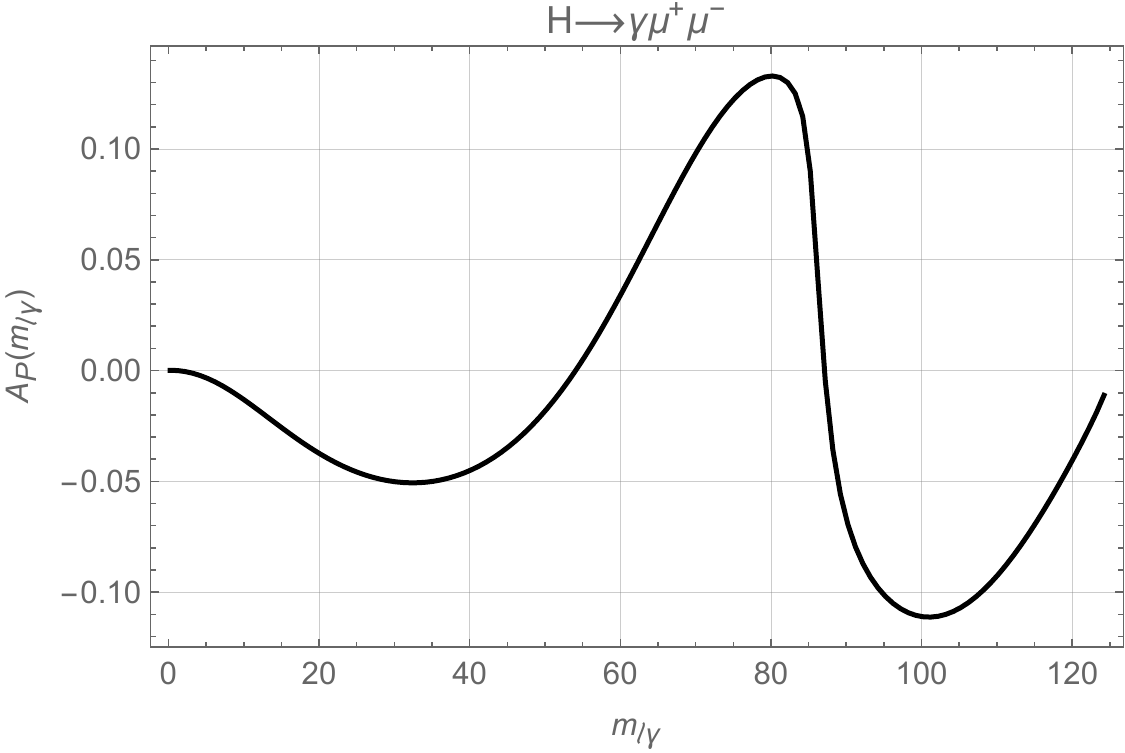}\\
(a)&(b)\\
\includegraphics[width=3in,height=2in]{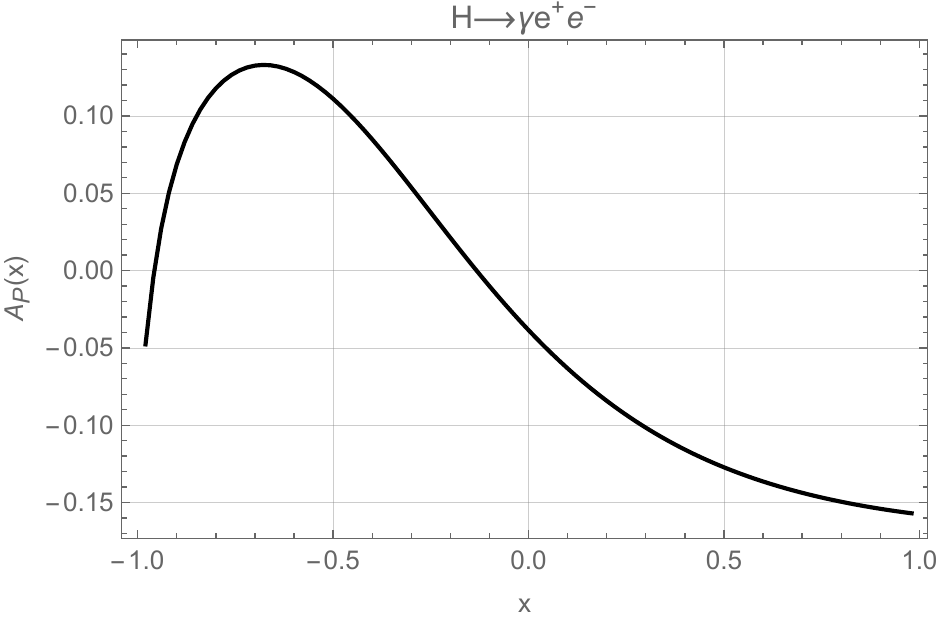}&
\includegraphics[width=3in,height=2in]{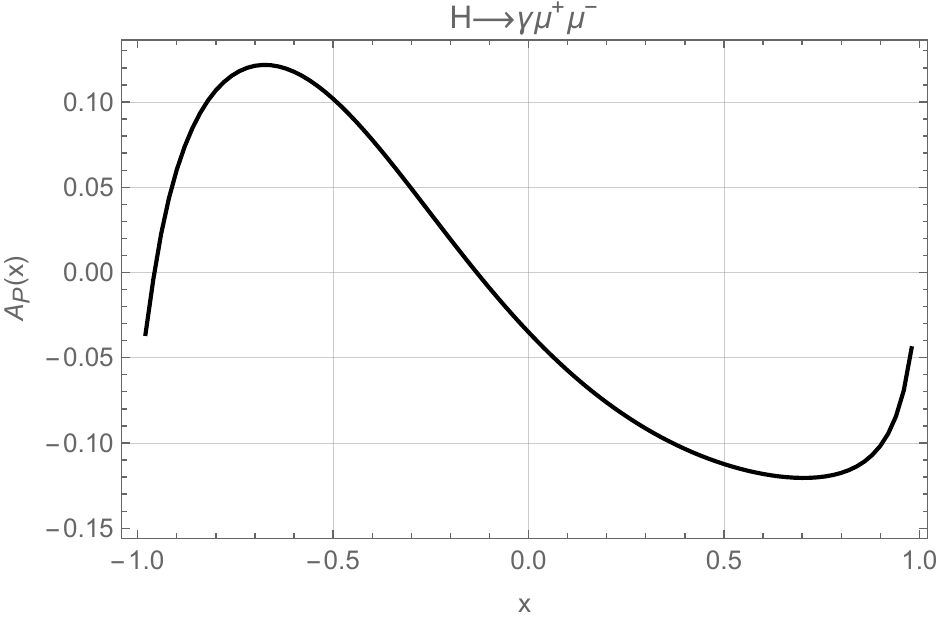}\\
(c)&(d)
\end{tabular}
\caption{Total photon polarization asymmetries $A_P$ against $x$ and $m_{\ell\gamma}$. (a) and (c) represent the total photon polarization asymmetries for electrons, while (b) and (d) represent muon case.}
\label{PAS}
\end{figure*}

Figs. \ref{PAS} (a) and (b) represent the photon polarization asymmetry, $A_P$, plotted against $m_{l\gamma}$ for the case of final-state electron and muon, respectively. We observe that the value of $A_P$ is positive for $55\lesssim m_{\ell\gamma} \lesssim85$ GeV, whereas it is negative for the rest of the kinematical region and the maximum value of $A_P$ reaches $\sim 0.14\text{ }(0.1)$ for the muon (electron) around $80$ GeV before the Z-resonance. 

Figs. \ref{PAS} (c) and (d) represent $A_P$ as a function of $x$ for the case of electron and muon, respectively. The value of $A_P$ increases smoothly from zero as $x$ starts increasing from -1 and reaches a maximum (minimum) value ($\pm 0.12$) near $x \approx \mp 0.7$ and becomes zero at $\theta \approx 0.54\pi$.

\begin{figure*}[htbp]
\centering
\begin{tabular}{ccc}
\includegraphics[width=3in,height=2in]{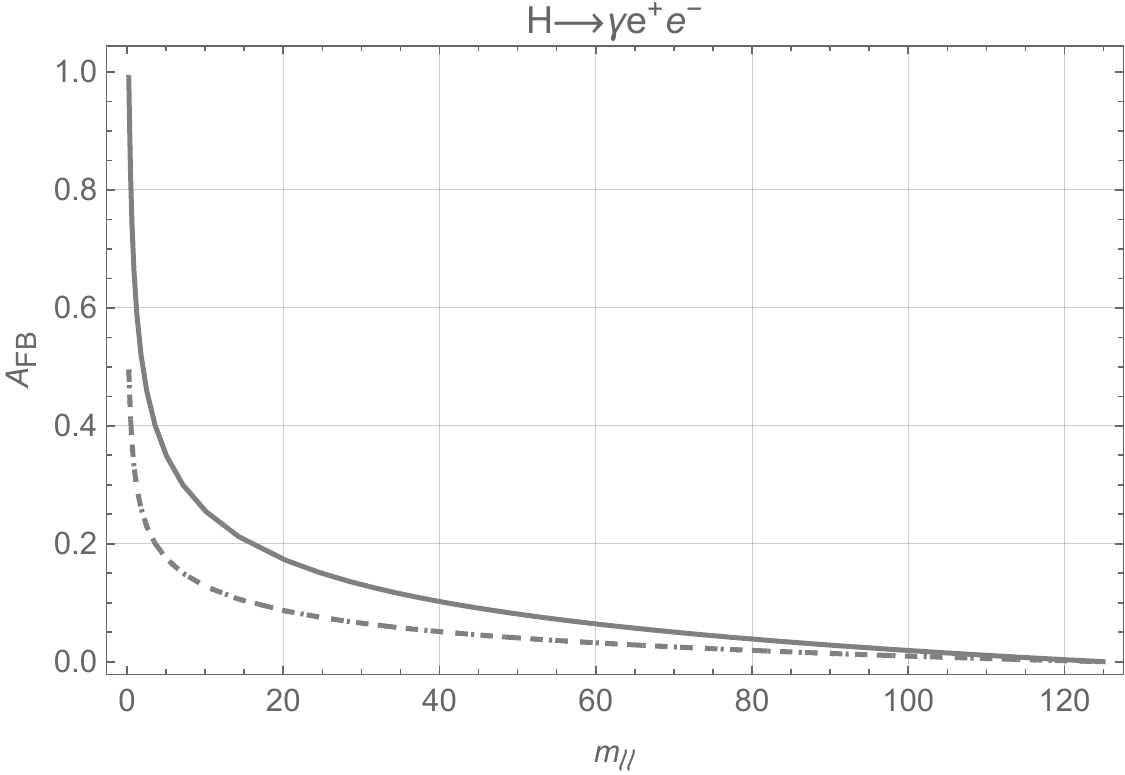}&
\includegraphics[width=3in,height=2in]{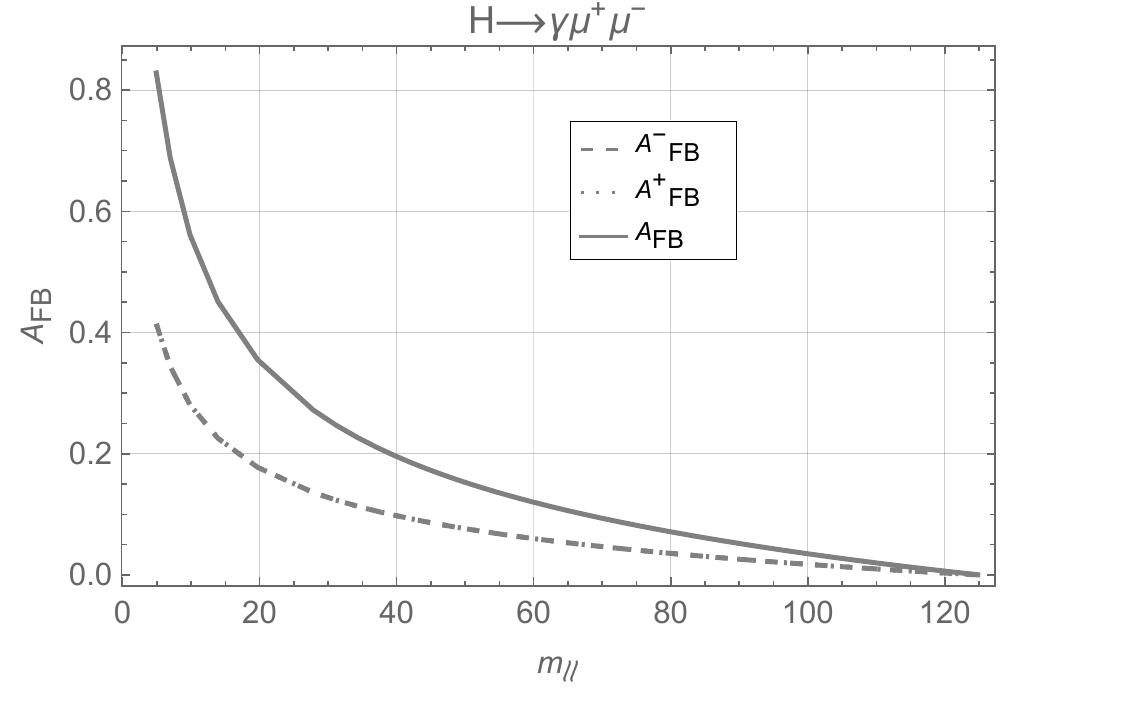}&\\
(a)&(b)
\end{tabular}
\caption{ Comparison of the tree level contribution to the unpolarized $A_{FB}$ (gray solid line) with the polarization effect on the tree level contribution to $A_{FB}^{(\pm)}$, when lepton or photon is polarized (dashed and dotted curves) where (a) and (b) show electron and muon cases respectively.}
\label{TreeFBA}
\end{figure*}

\begin{figure*}[htbp]
\centering
\begin{tabular}{ccc}
\includegraphics[width=3in,height=2in]{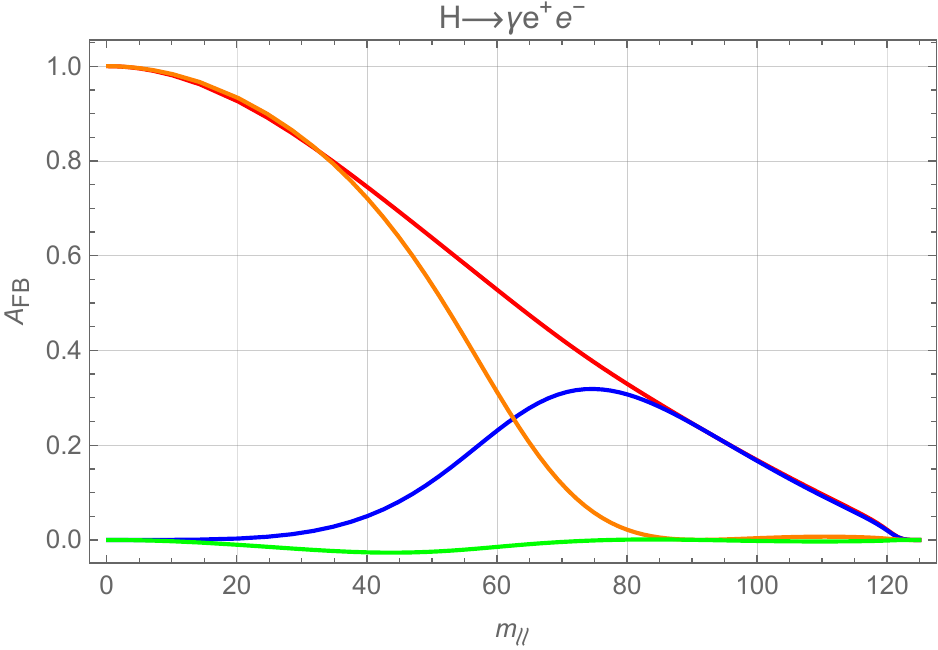}&
\includegraphics[width=3in,height=2in]{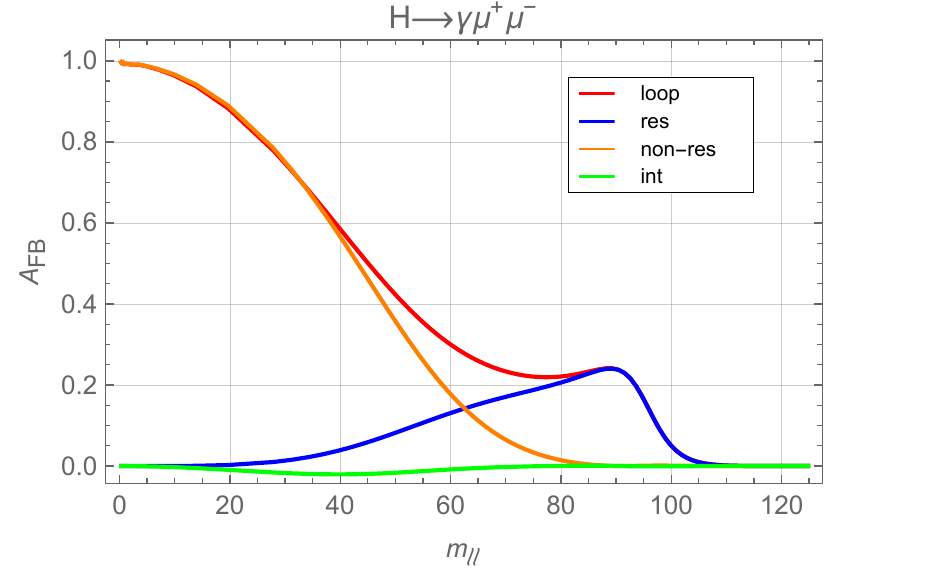}&\\
(a)&(b)
\end{tabular}
\caption{ The loop, the resonance, the non-resonance and the interference contributions of unpolarized forward-backward asymmetry $A_{FB}$ against $m_{\ell\ell}$ are represented by red, blue, orange and green curves, respectively. }
\label{FBAunpol}
\end{figure*}

\begin{figure*}[htbp]
\centering
\begin{tabular}{ccc}
\includegraphics[width=3in,height=2in]{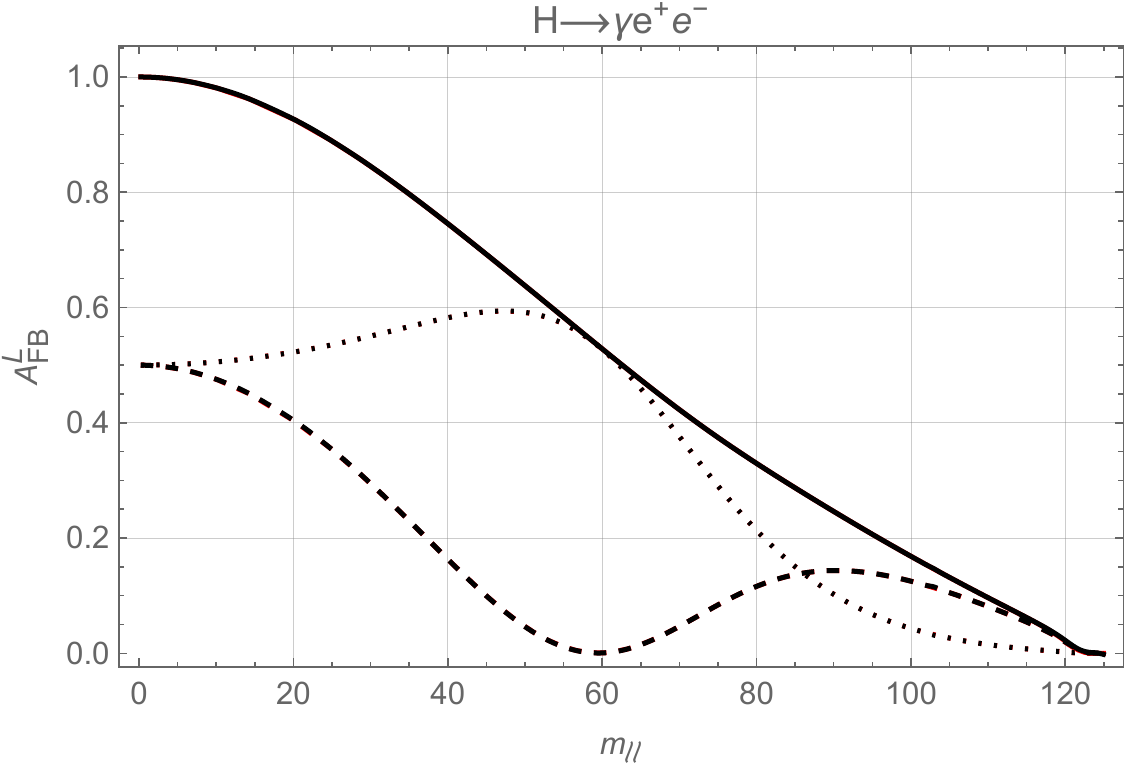}&
\includegraphics[width=3in,height=2in]{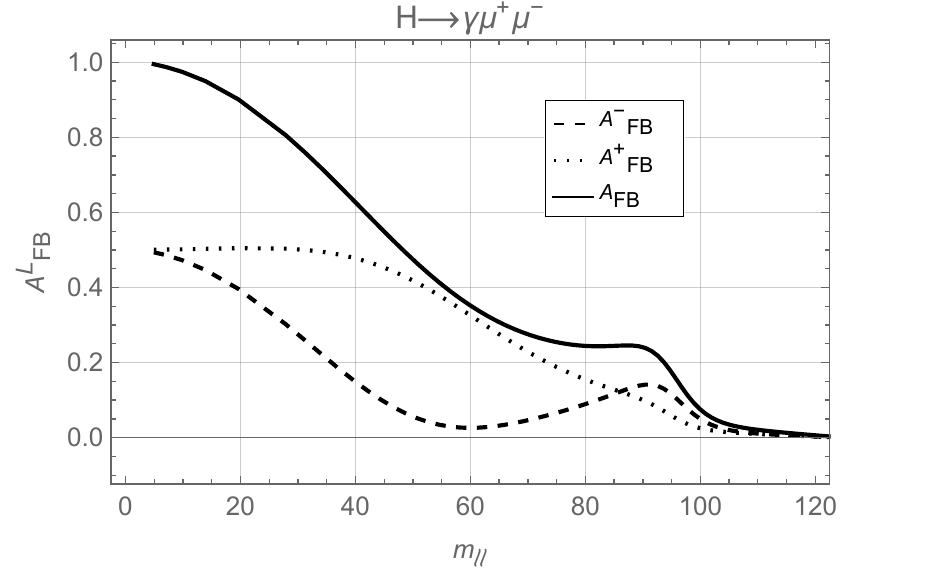}&\\
(a)&(b) \\
\includegraphics[width=3in,height=2in]{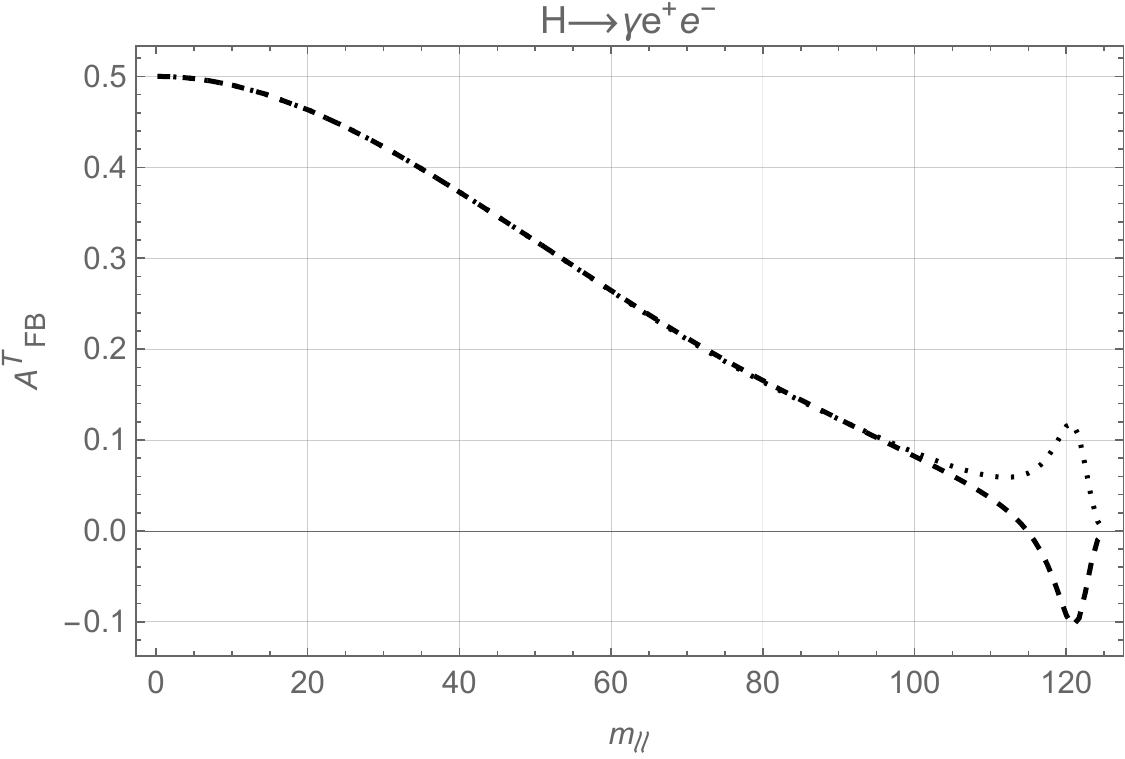}&
\includegraphics[width=3in,height=2in]{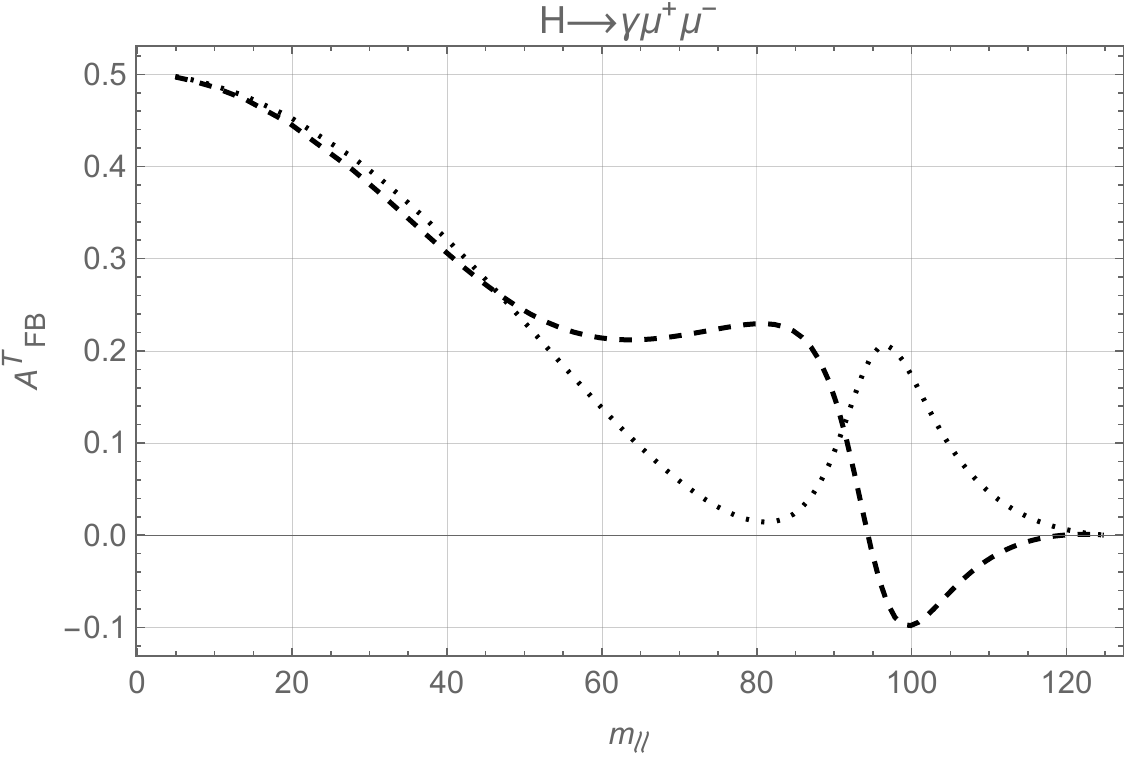}&\\
(c)&(d) \\
\includegraphics[width=3in,height=2in]{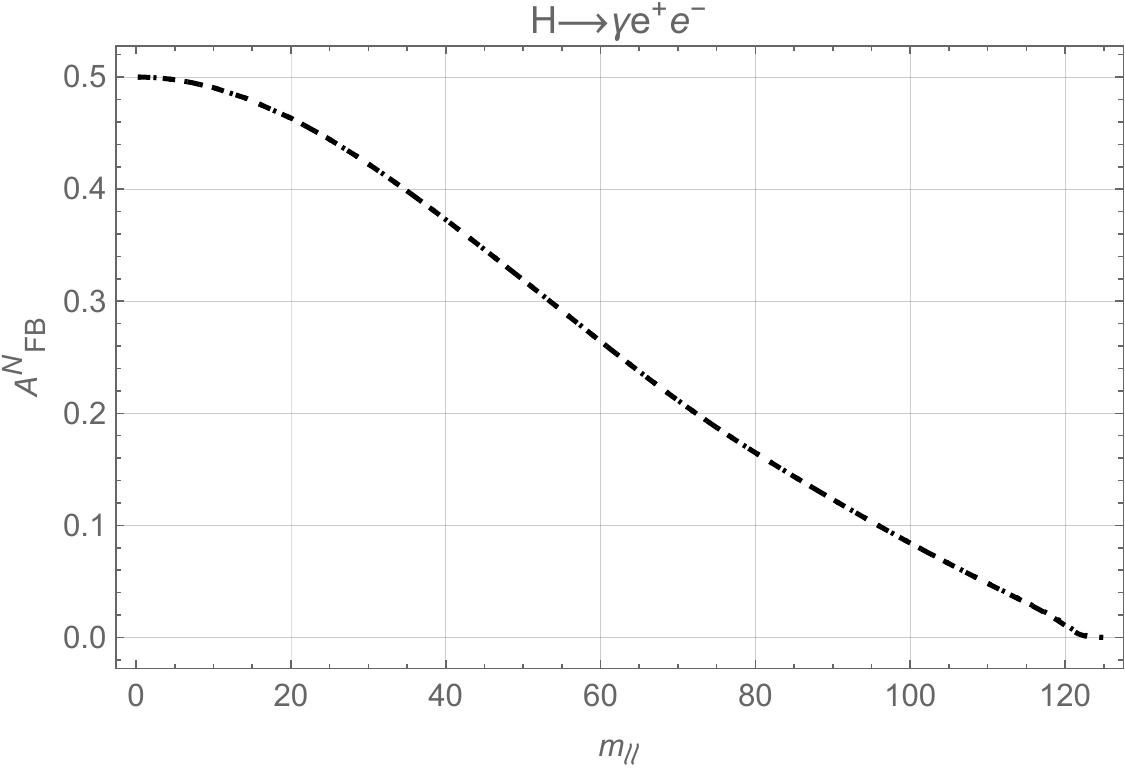}&
\includegraphics[width=3in,height=2in]{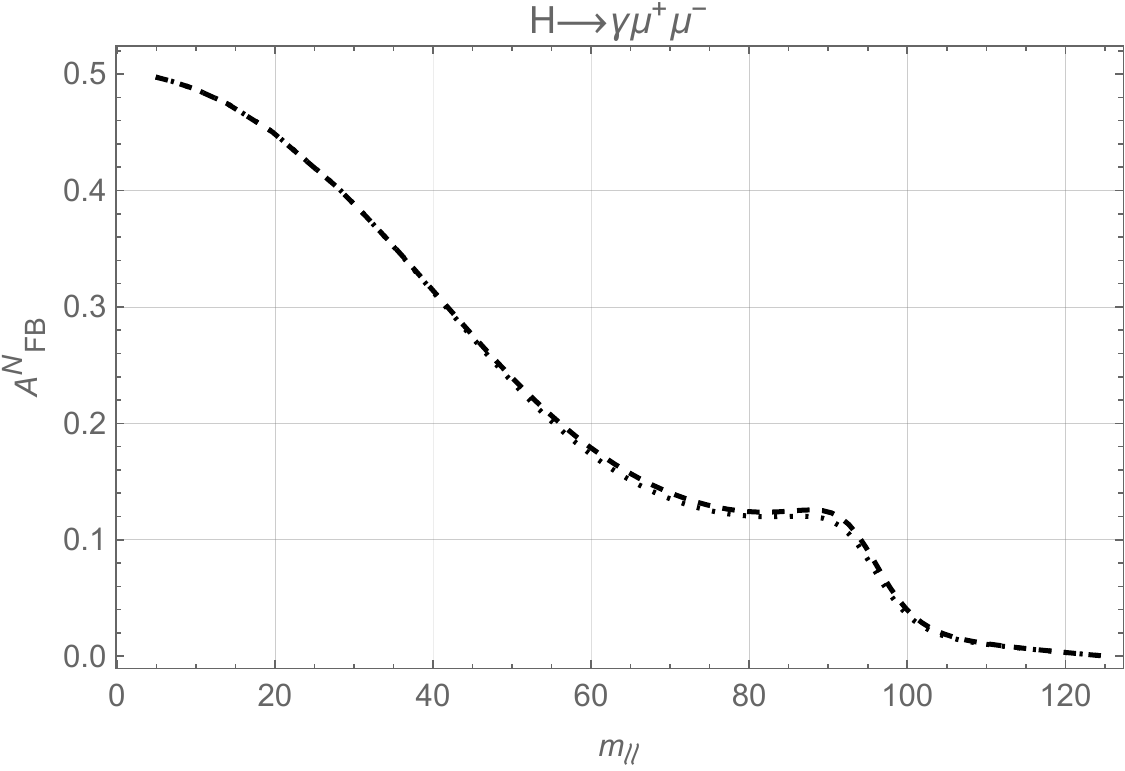}&\\
(e) & (f)
\end{tabular}
\caption{(a), (c) and (e) represent the longitudinally, transversely, and normally polarized forward-backward asymmetries $A_{FB}^{(i,\pm)}(s)$, respectively, for the electron, while (b), (d) and (f) are for the muon. Color scheme is same as defined in Fig. \ref{LPDRx}.}
\label{FBALP}
\end{figure*}

\begin{figure*}[htbp]
\centering
\begin{tabular}{ccc}
\includegraphics[width=3in,height=2in]{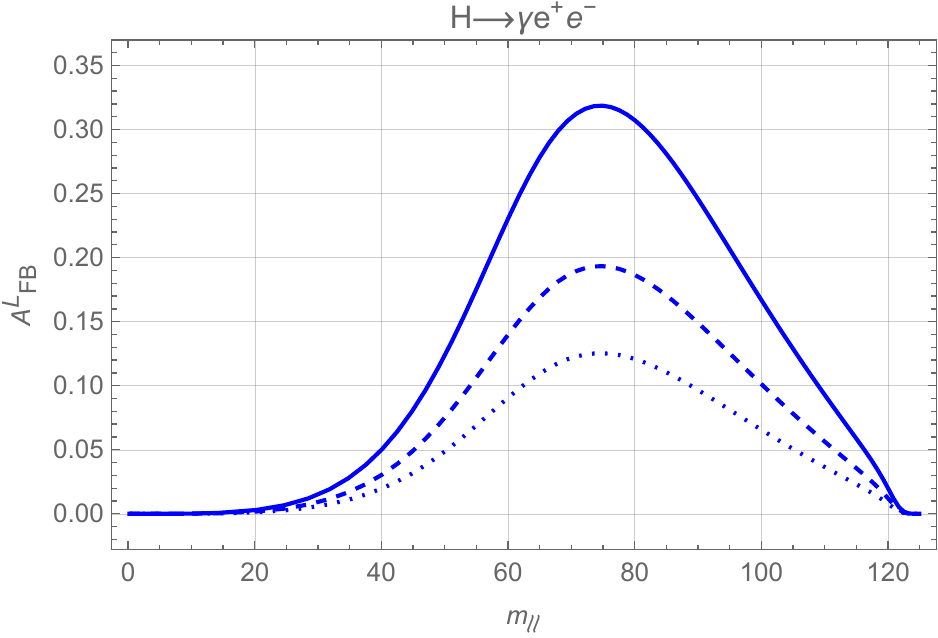}&
\includegraphics[width=3in,height=2in]{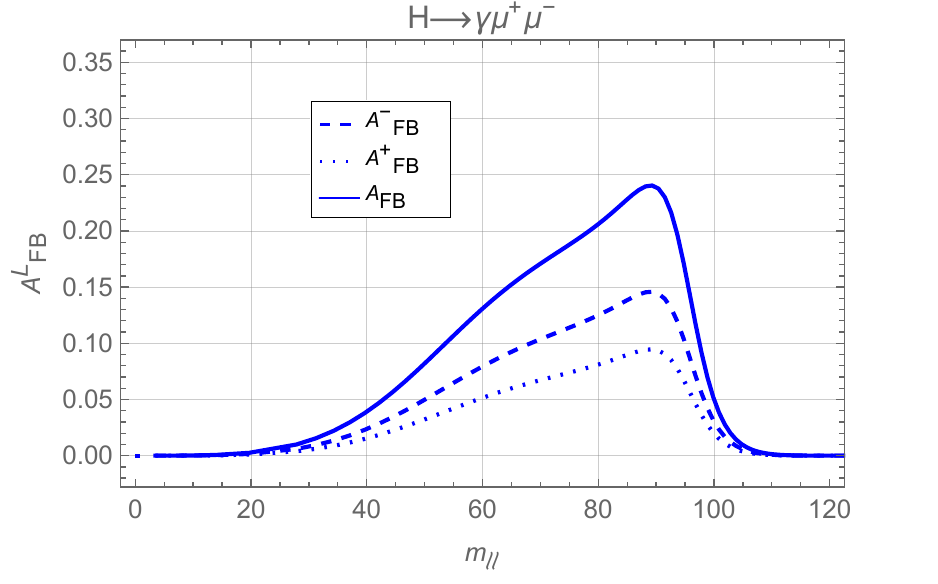}&\\
(a)&(b) \\
\includegraphics[width=3in,height=2in]{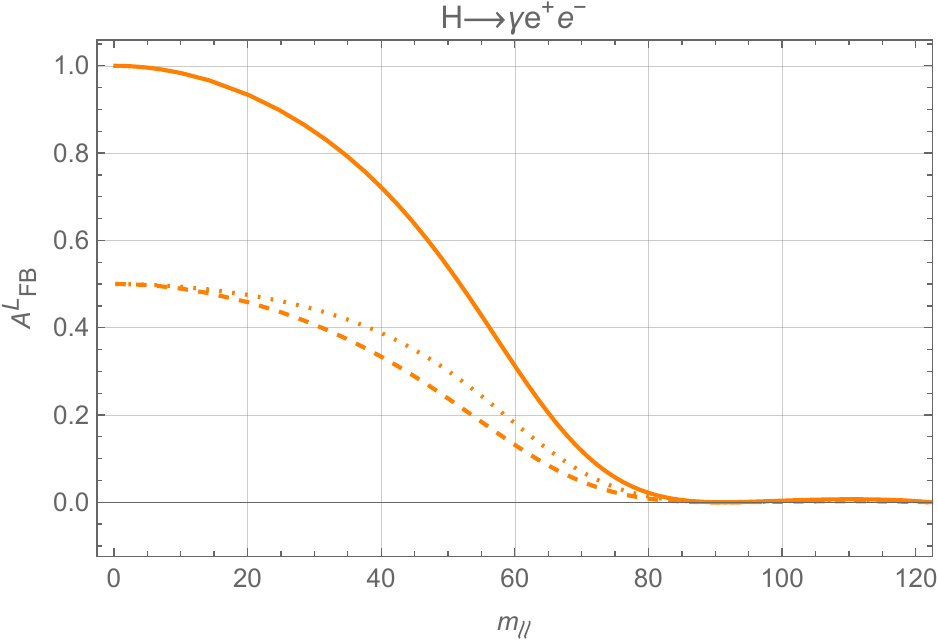}&
\includegraphics[width=3in,height=2in]{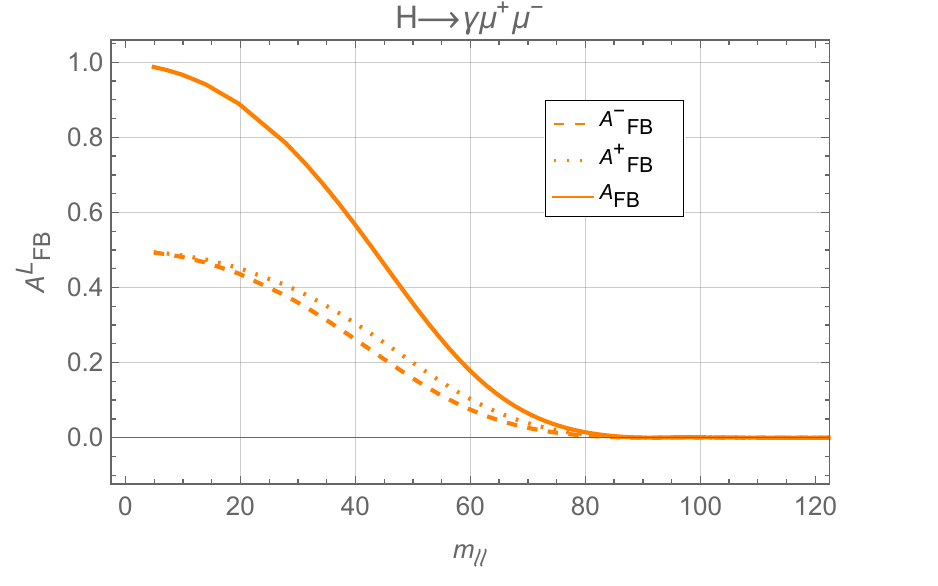}&\\
(c)&(d) \\
\includegraphics[width=3in,height=2in]{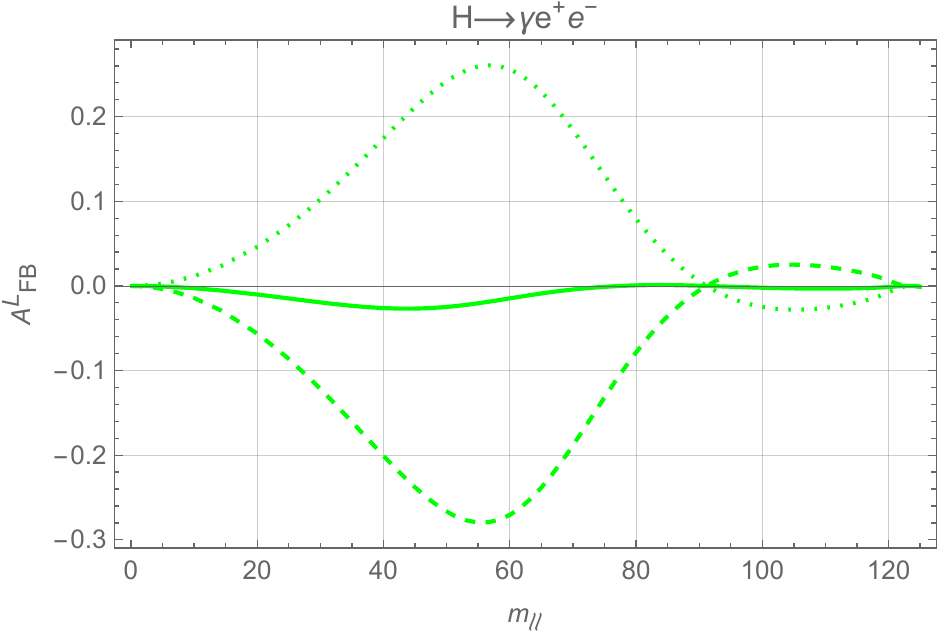}&
\includegraphics[width=3in,height=2in]{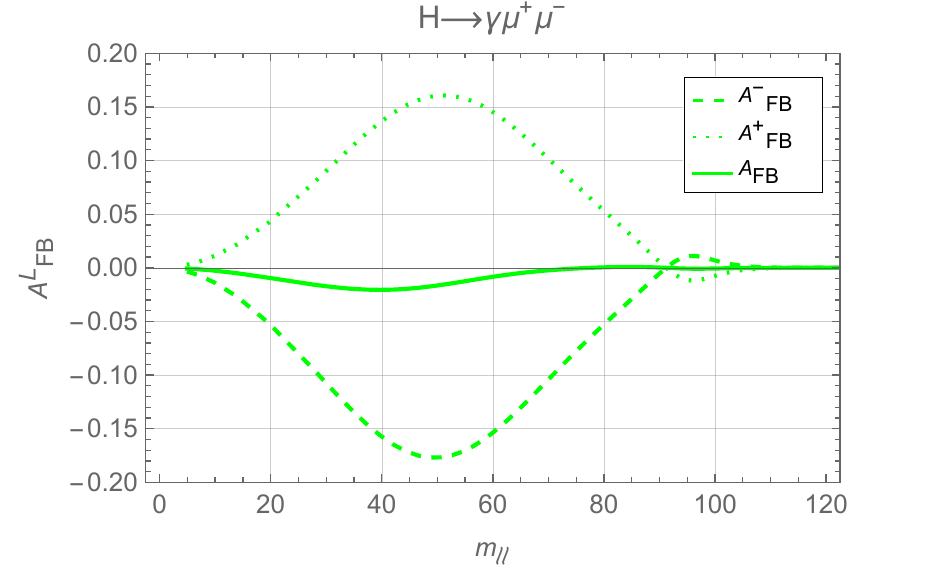}&\\
(e) & (f)
\end{tabular}
\caption{(a), (c) and (e) represent the resonance (blue curves), non-resonance (orange curves) and their interference (green curves) contributions for longitudinally polarized and unpolarized forward-backward asymmetries $A_{FB}^{(L,\pm)}(s)$, respectively, for the electron, while (b), (d) and (f) are for the muon. The legends of curves are same as defined in Fig.\ref{LPDRx}.}
\label{FBALrnr}
\end{figure*}

\subsection{Forward-Backward Asymmetry$A_{FB}$}
 As mentioned previously, there are no polarization effects  at tree level; therefore, the tree level $A_{FB}$ is also not sensitive to the polarization of the final particles. In Fig. \ref{TreeFBA} (a) and (b), we have shown $A_{FB}$ as a function of $m_{\ell\ell}$ where it can be seen that its value reaches maximum when $m_{\ell\ell}$ is minimum. With an increasing value of $m_{ll}$, there is a sharp decrease in the value of $A_{FB}$ ($\sim0.2$) up to $m_{ee}\sim20$ GeV and $m_{\mu\mu}\sim40$ GeV. However, after these values of $m_{\ell\ell}$, $A_{FB}$  gradually decreases and becomes zero at the maximum value of $m_{ll}$. 

In Fig. \ref{FBAunpol}, we have shown the unpolarized loop level and the total $A_{FB}$. Here, it is observed that after $m_{\ell\ell}=60$ GeV to the end, 
the tree level of muon is comparable to the loop contribution, except in the Z-resonance region. Therefore, the asymmetry of muon is smaller than that of electron excluding the resonance region, the value of $A_{FB}$ for the muon case is suppressed in comparison to the case of the electron which creates concavity in front of the resonance region for the case of muon. This difference is due to the fact that the tree- and loop-level contributions of the decay rates for the case of $\mu$ are comparable, therefore, the denominator in the expression of $A_{FB}$ (see Eq. \ref{AFBformula}) for the muon is larger than for the electron, which causes this suppression resulting in concavity. This result is consistent with the finding of ref. \cite{Kachanovich:2020xyg} . Consequently, in this region, $A_{FB}$ becomes flavor dependent, which may also be a good complementary observable to check the Yukawa coupling.

\subsubsection{Lepton-Polarized Forward-Backward Asymmetry $A_{FB}^{(i,\pm)}$}
 In Figs. \ref{FBALP} (a) and (b), depict the behavior of the total unpolarized and polarized $A_{FB}$. and the polarized when the final state electron and muon are negatively and positively longitudinally polarized $A_{FB}^{(L,\pm)}$.Figs. \ref{FBALP} (a) and (b), depict the behavior of $A_{FB}^{(i,\pm)}$ when the final state lepton is longitudinally polarized. The total unpolarized $A_{FB}$ is depicted with a solid black curve.
Here, one can see that the unpolarized $A_{FB}$ is a decreasing function throughout the $m_{\ell\ell}$ region. Although for the case of positive longitudinally polarized electron, $A_{FB}^{(L,+)}$ is slightly an increasing function of $m_{\ell\ell}$ up to $\sim60$ GeV while for the case of muon, the value of $A_{FB}^{(L,+)}$ remains the same up to $\sim40$ GeV.  After these values of $m_{\ell\ell}$, $A_{FB}^{(L,+)}$ both for the electron and the muon decreases and becomes zero at the maximum value of $m_{ll}$. 

However, the negatively polarized asymmetry $A_{FB}^{(L,-)}$ is a decreasing function of $m_{ll}$ and reaches a minimum value around $m_{ll}$ $\sim60$ GeV, for the electron and muon cases. The difference in the values of asymmetries for negative and positive polarized final-state lepton is maximum around the region $m_{\ell\ell}$ $\sim60$ GeV. To explain the difference in the values of $A_{FB}^{(L,+)}$ and $A_{FB}^{(L,-)}$(see Figs. \ref{FBALP} (a,b)), we have shown the resonance, non-resonance and interference contributions of $A_{FB}^{(L,\pm)}$ in Fig. \ref{FBALrnr}. The blue (orange) solid, dotted and dashed curves show resonance (non-resonance) contributions while green curves show their interference terms. One can easily observe that the interference term accounts for the difference in $A_{FB}^{(L,+)}$ and $A_{FB}^{(L,-)}$ around 50 GeV.

In Figs. \ref{FBALP}  (c) and (d), we have shown $A_{FB}$ for the case of transversely polarized final-state leptons $A_{FB}^{(T,\pm)}$. For an electron, there are no polarization effects except near the maximum value of $m_{ee}$, where  $A_{FB}^{(T,-)}$becomes negative and  $A_{FB}^{(T,+)}$ becomes positive with almost the same magnitude, and at the maximum value of $m_{ee}$, the values of both asymmetries become zero. For the muon, $A_{FB}^{(T,\pm)}$ is not sensitive to the polarization up to  $m_{\mu\mu}\sim50$GeV; after that, the polarized effects appear, and due to this, the value of $A_{FB}^{(T,+)}$ decreases and $A_{FB}^{(T,-)}$ almost remains the same. Around the resonance, the $A_{FB}^{(T,+)}$ becomes dominant and $A_{FB}^{(T,-)}$ becomes small and after the resonance, we observe the zero crossing in $A_{FB}^{(T,-)}$. Around $100$ GeV, the $A_{FB}^{(T,-)}(A_{FB}^{(T,+)})$ observes a local dip (tip) and at the maximum values of $m_{ee}$, both of these asymmetries become zero. A similar behavior was also reported in ref. \cite{shahkar} for the positive and negatively polarized decay rate. 
The $A_{FB}^{(N,\pm)}$ is shown in Figs. \ref{FBALP} (e) and (f) for electron and muon respectively. There is almost no effect of lepton polarization on the $A_{FB}^{(N,\pm)}$.
It is worth mentioning that similar to unpolarized $A_{FB}$, the concavities also appear around the Z-resonance in the polarized $A_{FB}^{(\pm)}$ for the case of muon. Again this 
behavior arises due to the tree and loop level contributions becoming comparable for the case of muon, causing the suppression in the form of concavities.

\subsubsection{Photon-Polarized Forward-Backward Asymmetry $A_{FB}^\pm$}
\begin{figure*}[htbp]
\centering
\begin{tabular}{cc}
\includegraphics[width=3in,height=2in]{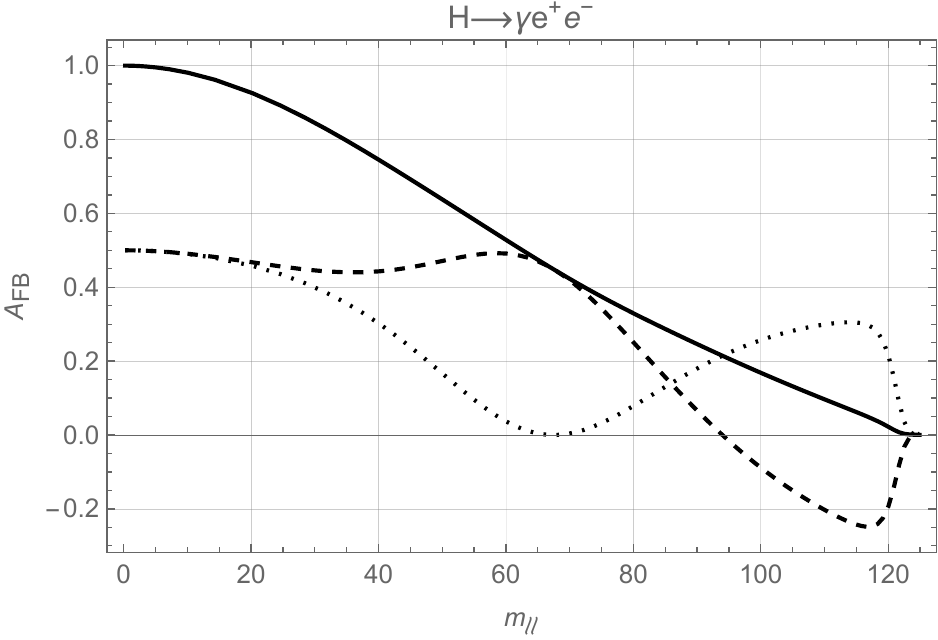}&
\includegraphics[width=3in,height=2in]{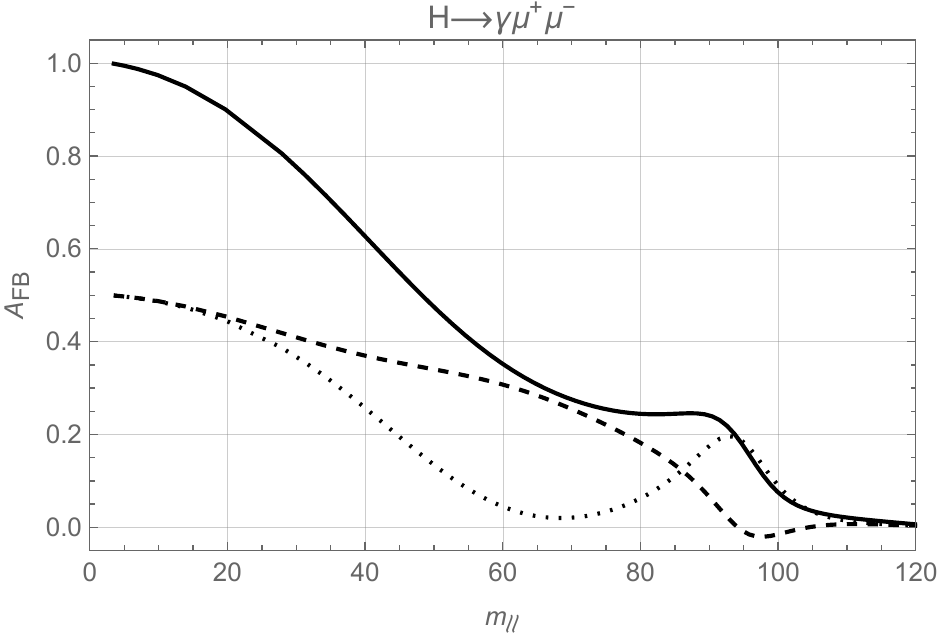}\\
(a)&(b)\\
\includegraphics[width=3in,height=2in]{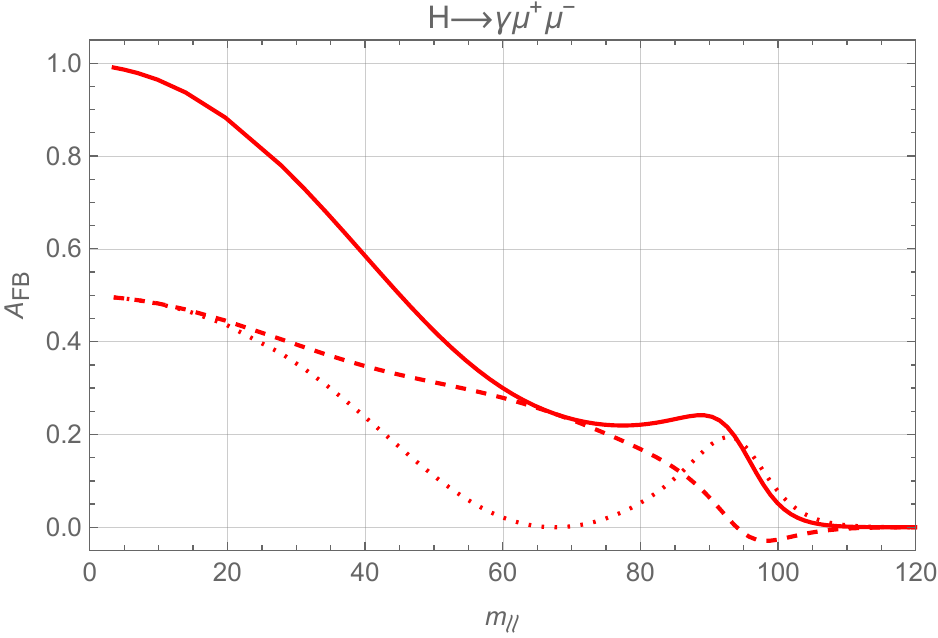}&
\includegraphics[width=3in,height=2in]{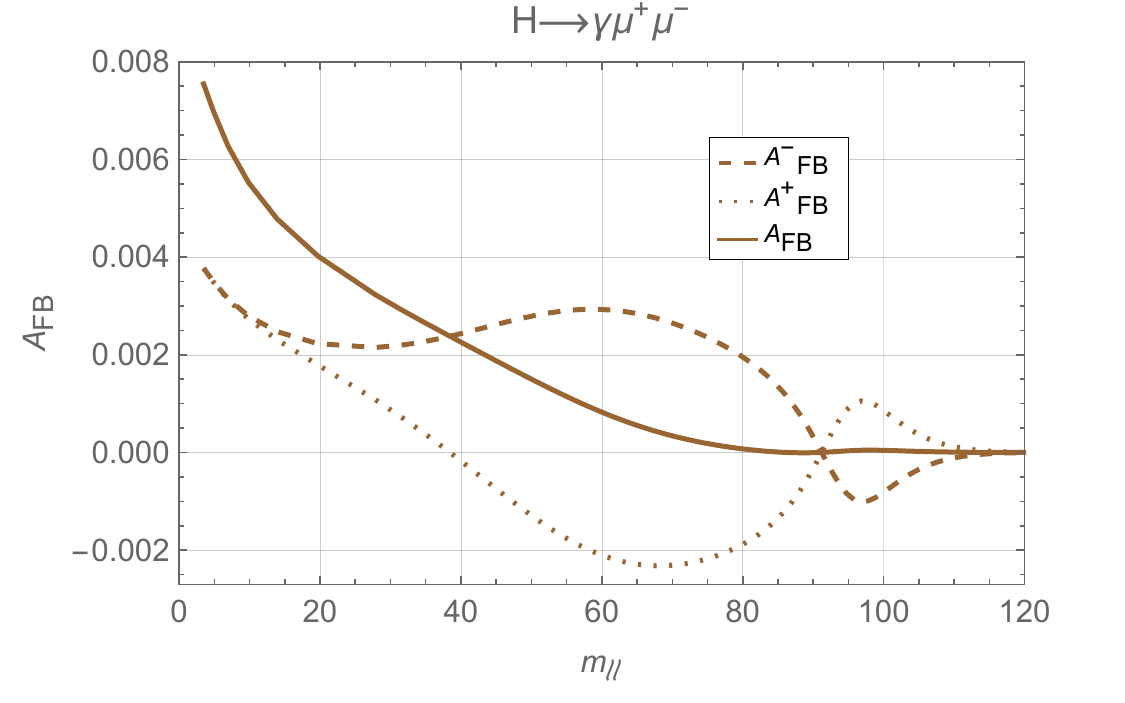}\\
(c)&(d)
\end{tabular}
\caption{ (a) and (b) show the total contribution for the final state electrons and muons, respectively, for the photon polarized forward-backward asymmetry against $m_{\ell\ell}$ while (c) and (d) show the loop and tree-loop interference contributions for the case of final state muons. Color scheme and legends of curves are same as defined in Fig. \ref{LPDRx}.}
\label{FBAGP}
\end{figure*}

\begin{figure*}[]
\centering
\begin{tabular}{ccc}
\includegraphics[width=3in,height=2in]{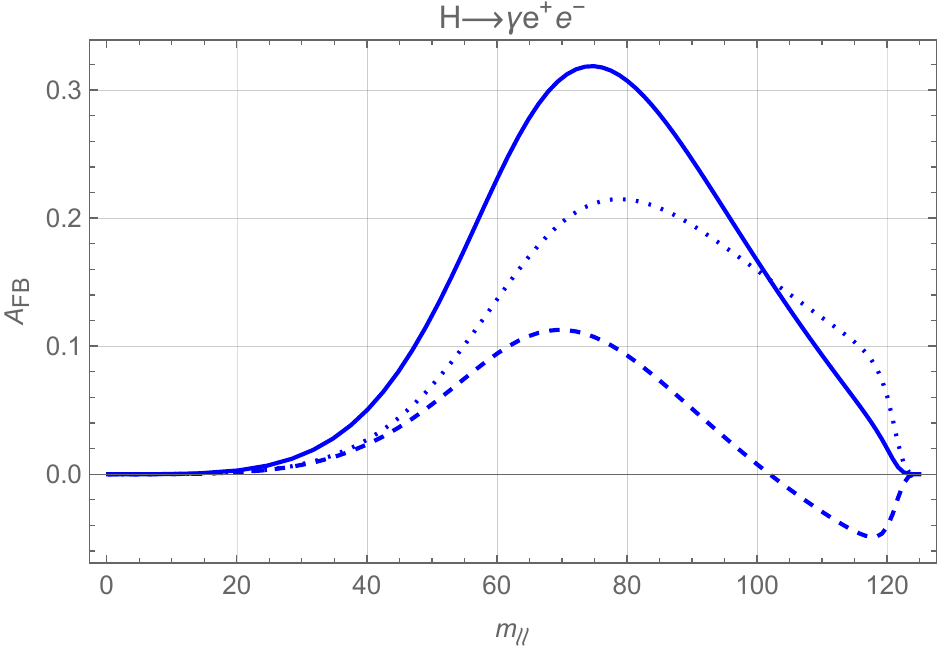}&
\includegraphics[width=3in,height=2in]{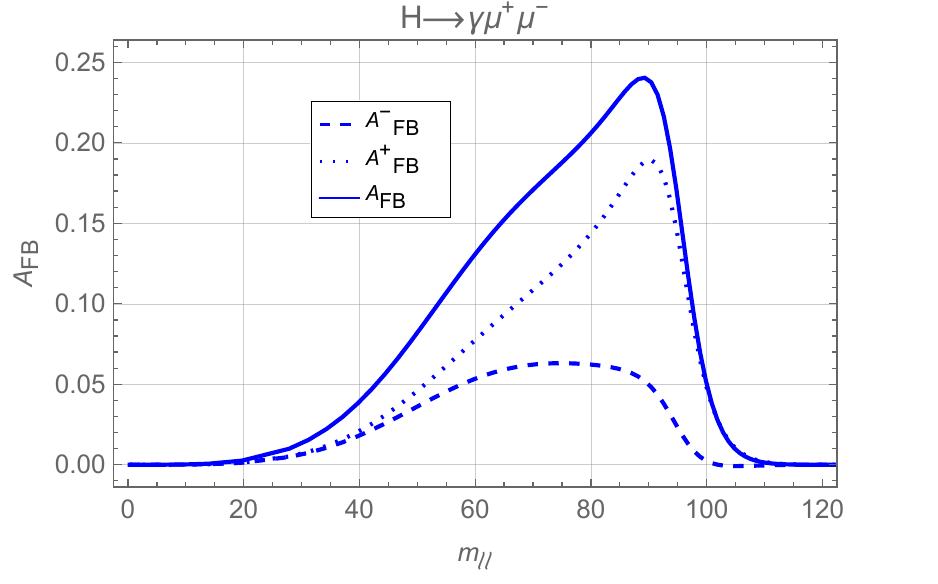}&\\
(a)&(b) \\
\includegraphics[width=3in,height=2in]{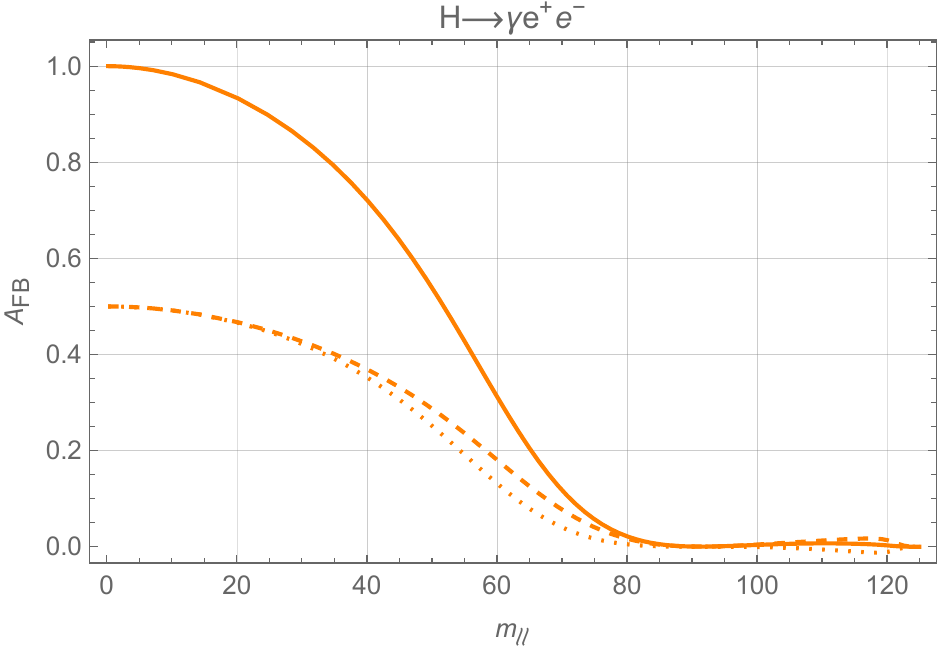}&
\includegraphics[width=3in,height=2in]{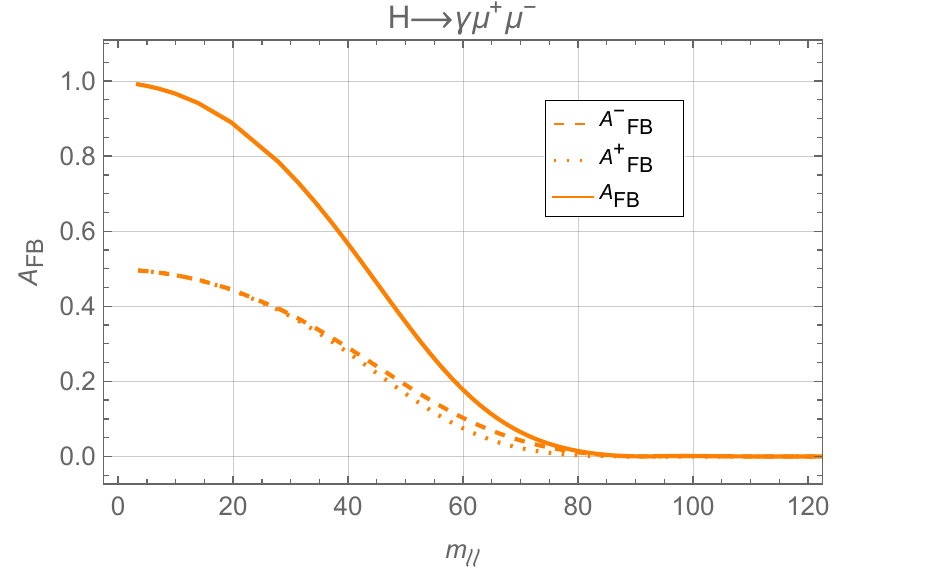}&\\
(c)&(d) \\
\includegraphics[width=3in,height=2in]{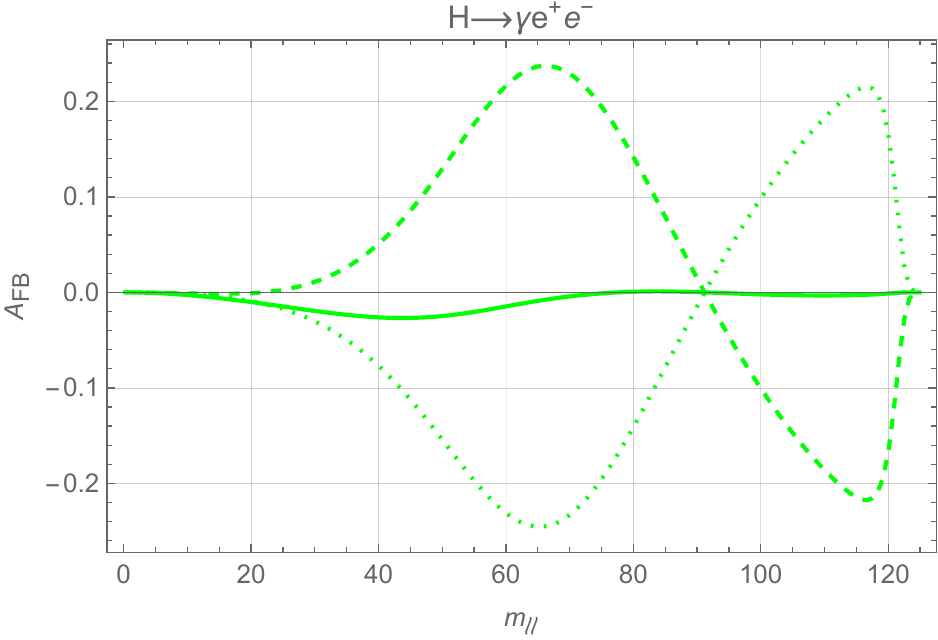}&
\includegraphics[width=3in,height=2in]{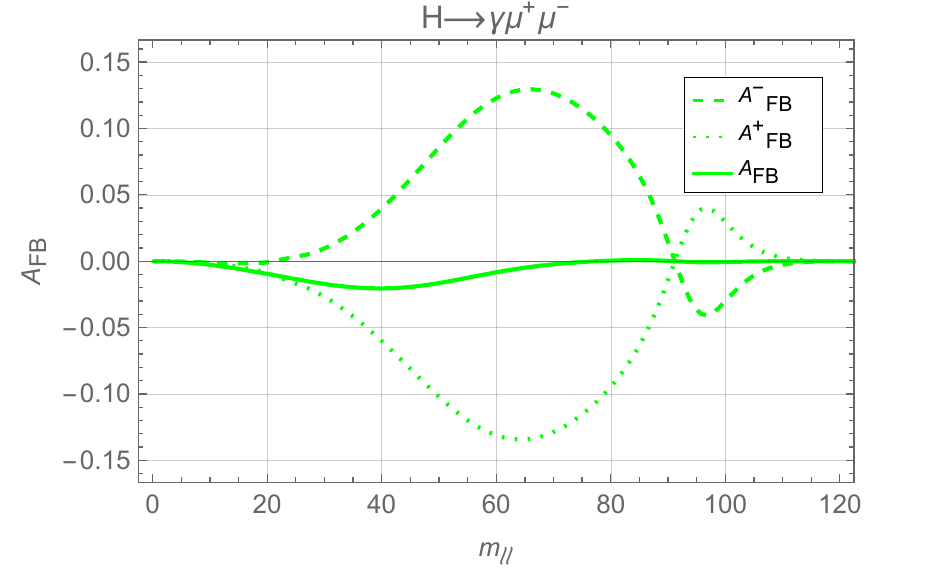}&\\
(e) & (f)
\end{tabular}
\caption{(a), (c) and (e) represent the resonance (blue curves), non-resonance (orange curves) and their interference (green curves) contributions for photon-polarized and unpolarized forward-backward asymmetries $A_{FB}^{(\pm)}(s)$, respectively, for the electron, while (b), (d) and (f) are for the muon. The legends of curves are same as defined in Fig.\ref{LPDRx}.}
\label{GPFBArnr}
\end{figure*}

Fig. \ref{FBAGP} represents the $A_{FB}^\pm$ when the final state photon is polarized. It can be seen from these plots that the value of $A_{FB}^{+}$ is positive for electron and muon throughout kinematical region while the value of $A_{FB}^{-}$ is positive (negative) before (after) the resonance for the electron case. For muon, $A_{FB}^{-}$ remains positive throughout the kinematical region. Similar to the case of $A_{FB}^{L,\pm}$, there is also the maximum difference in the values of $A_{FB}^{\pm}$ around 60 GeV. We investigate this behavior in the same manner as discussed for the case of $A_{FB}^{(L,\pm)}$ and have drawn the resonance, non-resonance and their interference contributions of $A_{FB}^{\pm}$ in Fig. \ref{GPFBArnr}. Here, the interference contribution is again responsible for the difference in the values of $A_{FB}^{+}$ and $A_{FB}^{-}$ around $70$ GeV.

For the same reason as above, notice in Fig. \ref{FBAGP}(b) that we see the similar concavities around the resonance in the plots for photon polarized $A_{FB}^{\pm}$ for final state muons. 


\begin{figure*}[htbp]
\centering
\begin{tabular}{cc}
\includegraphics[width=3in,height=2.5in]{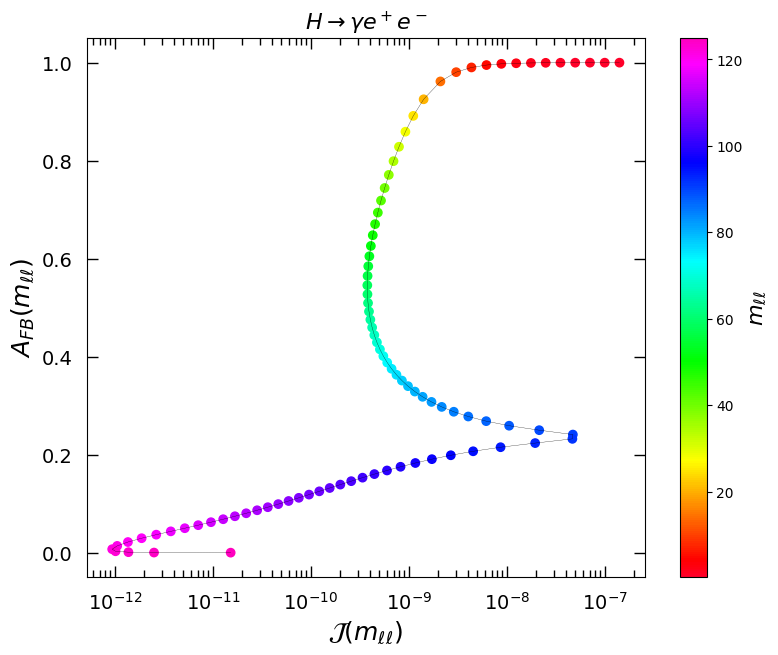}&
\includegraphics[width=3in,height=2.5in]{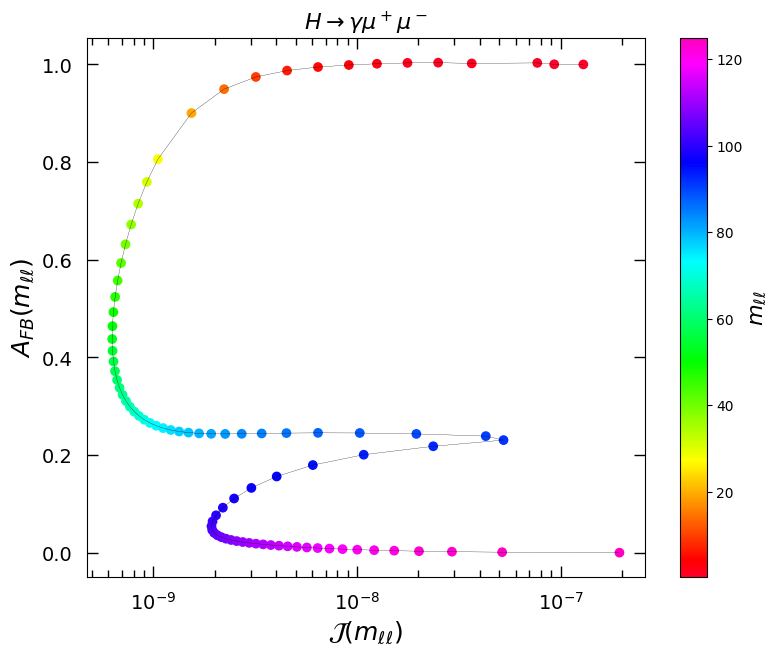}\\
(a)&(b)\\
\includegraphics[width=3in,height=2.5in]{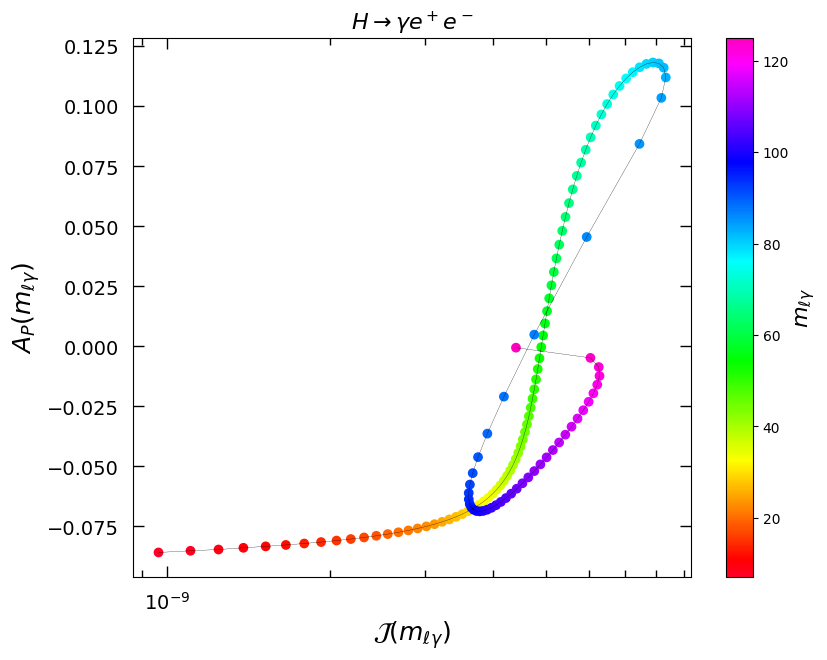}&
\includegraphics[width=3in,height=2.5in]{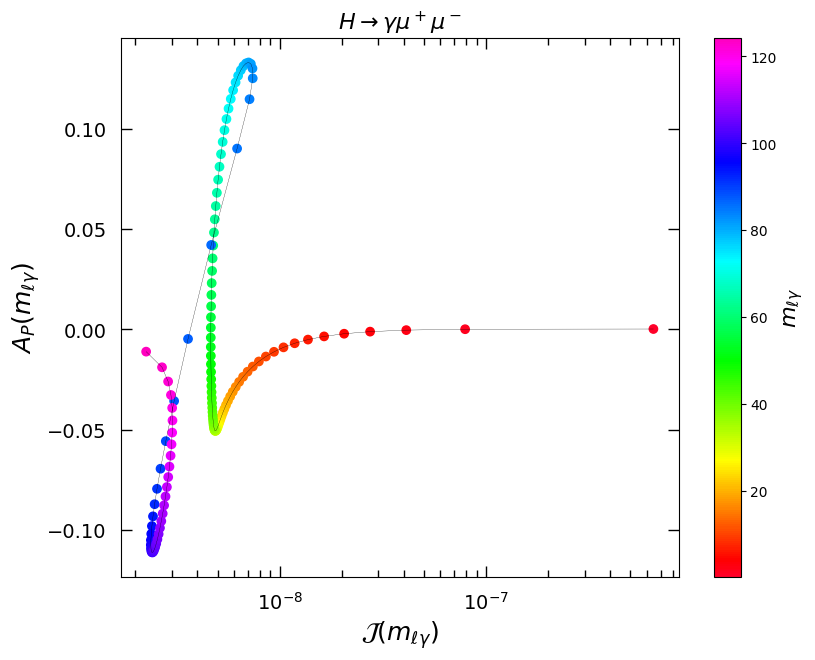}\\
(c)&(d)
\end{tabular}
\caption{(a) and (b) are the correlations between the $A_{FB}(m_{\ell\ell})$ and the $\mathcal{J}(m_{\ell\ell})$ for electron and muon respectively, while (c) and (d) are the correlations between the $A_P(m_{\ell\gamma})$ and the $\mathcal{J}(m_{\ell\gamma})$ for electron and muon respectively.}
\label{corr}
\end{figure*}

\subsection{Correlation between Decay Rates $\mathcal{J}$ and Asymmetries $A_{FB},A_P$ }
In Figs. \ref{corr}  (a,b) and (c,d), we have shown the correlation between the $\mathcal{J}$ and $A_{FB}$  and $\mathcal{J}$ and $A_{P}$, respectively, for both the electron and muon cases. In Fig. \ref{corr} (a, b), one can see that for the case of electrons at low values of $m_{\ell\ell}$ and after the resonance region there is direct or one to one correlation between $\mathcal{J}$ and $A_{FB}$, while for the case of muons, there is no single-valued correlation region between $\mathcal{J}$ and $A_{FB}$. Therefore, precise measurement of the decay rates of $H\to e^+e^-\gamma $  at low values of $m_{\ell\ell}$ and after the resonance one can predict the value of  $A_{FB}$ for the electron in these regions. Similarly, from (c, d), it is observed that the correlation between $\mathcal{J}$ and $A_P$, below the $m_{\ell\ell}\lesssim20$ GeV, is also single-valued for both the electron and muon cases, consequently, the precise measurement of  $H\to \ell^+\ell^-\gamma $ ($\ell=e,\mu$)  can give the prediction of the value of $A_{P}$ in this region.

Furthermore, the numerical values of the decay rates and the asymmetries in different $q^2$ bins are tabulated in Table \ref{wc table}. 

\begin{table*}
\begin{tabular}{c|c|c|c|c|c|cc}
\hline\hline
$m_{\ell\ell}  $&$10-30$&$30-50$&$50-70$&$70-90$&$90-110$& Total
\\ \hline
\ \  $10^{9}\times\Gamma^{-}_{H\to e^{+}e^{-}\gamma}$ \ \  & \ \  $10.2$ \ \  & \ \  $2.53$ \ \  & \ \ $0.285$ \ \  & \ \  $31.9$ \ \  & \ \  $90.1$ \ \  & \ \ $133.2$ \ \    \\  
  \hline
  \ \  $10^{9}\times\Gamma^{+}_{H\to e^{+}e^{-}\gamma}$ \ \  & \ \  $13.1$ \ \  & \ \  $8.4$ \ \  & \ \ $7.47$ \ \  & \ \  $31.7$ \ \  & \ \  $54.8$ \ \  & \ \ $113.3$ \ \    \\  
\hline
\ \  $10^{9}\times\Gamma^{-}_{H\to \mu^{+}\mu^{-}\gamma}$ \ \  & \ \  $10.5$ \ \  & \ \  $3.21$ \ \  & \ \ $1.53$ \ \  & \ \  $34.3$ \ \  & \ \  $96.4$ \ \  & \ \ $154.7$ \ \    \\  
  \hline
  \ \  $10^{9}\times\Gamma^{+}_{H\to \mu^{+}\mu^{-}\gamma}$ \ \  & \ \  $13.4$ \ \  & \ \  $91.2$ \ \  & \ \ $8.75$ \ \  & \ \  $34.1$ \ \  & \ \  $61.2$ \ \  & \ \ $135.1$ \ \    \\  
  \hline
  \ \  $A_{FB}(m_{ee})$ \ \  & \ \  92.23\% \ \  & \ \  74.43\% \ \  & \ \ 52.90\% \ \  & \ \  33.13\% \ \  & \ \  16.94\% \ \  & \ \  51.98\% \ \    \\ \hline
 \ \  $A_{FB}(m_{\mu\mu})$ \ \  & \ \  89.0\% \ \  & \ \  62.65\% \ \  & \ \ 35.97\% \ \  & \ \  25.0\% \ \  & \ \  10.36\% \ \  & \ \  44.06\% \ \    \\ \hline
  \ \  $A_P(m_{ee})$ \ \  & \ \  -7.68\% \ \  & \ \  -4.72\% \ \  & \ \ 3.23\% \ \  & \ \  8.15\% \ \  & \ \  6.27\% \ \  & \ \  -2.27\% \ \    \\ \hline
  \ \  $A_P(m_{\mu\mu})$ \ \  & \ \  -3.53\% \ \  & \ \  -4.15\% \ \  & \ \ 3.61\% \ \  & \ \  9.07\% \ \  & \ \  10.15\% \ \  & \ \  -1.51\% \ \    \\ \hline
  \ \  $A^{L+}_{FB}(m_{ee})$ \ \  & \ \  46.8\% \ \  & \ \  44.8\% \ \  & \ \ 47.5\% \ \  & \ \  24.8\% \ \  & \ \  -8.1\% \ \  & \ \ 27.0\% \ \    \\  \hline
  \ \  $A^{L-}_{FB}(m_{ee})$ \ \  & \ \  45.4\% \ \  & \ \  29.6\% \ \  & \ \ 5.4\% \ \  & \ \  8.3\% \ \  & \ \  25.1\% \ \  & \ \ 25.0\% \ \    \\  \hline
  \ \  $A^{L+}_{FB}(m_{\mu\mu})$ \ \  & \ \  45.2\% \ \  & \ \  37.2\% \ \  & \ \ 30.4\% \ \  & \ \  17.5\% \ \  & \ \  0.2\% \ \  & \ \ 23.8\% \ \    \\  \hline
  \ \  $A^{L-}_{FB}(m_{\mu\mu})$ \ \  & \ \  43.8\% \ \  & \ \  25.5\% \ \  & \ \ 5.6\% \ \  & \ \  7.5\% \ \  & \ \  10.2\% \ \  & \ \ 20.0\% \ \    \\  
\hline
\hline
\end{tabular}
\caption{The decay rates, forward-backward (photon polarization) asymmetry $A_{FB}$  ($A_{P}$) in diffferent bins of $m_{\ell\ell}$  for $H\to\ell^+\ell^-\gamma$ decays. } \label{wc table}
\end{table*}

\section{Conclusion}\label{con}
 In this work, we have calculated the decay rates of $H\to\ell^+\ell^-\gamma$ ($\ell=e,\mu$), forward-backward asymmetries $A_{FB}^{\pm}$  and the polarization asymmetries $A_{P}^{\pm}$  for the final state polarized photon. In addition, the $A_{FB}^{(i,\pm)}$ is also calculated for the polarized final state lepton. We found that these observables, particularly, the mentioned asymmetries, can provide valuable insights about different features such as the Yukawa couplings, as well as the distinction between resonant and non-resonant contributions to the process $H\to\ell^+\ell^-\gamma$. Moreover, the asymmetries studied in this work may provide a helpful tool in ameleorating the discrepancies in the decay rates of $H\to e^+e^-\gamma$. Our analysis presents a detailed framework for assessing polarization-dependent observables in this decay process. Furthermore, by separately examining the polarization effects of the final-state photon and lepton on $A_{FB}$, we highlight the crucial role of loop-induced contributions in shaping the observable outcomes. In contrast to the unpolarized case, where interference effects are typically negligible, for the polarized case, interference terms often play an important role. We have also observed that the polarization-dependent observables serve as a robust tool for probing both the resonant and non-resonant characteristics of this decay channel. 
 
 In addition, it is known that in the loop-level process, the new particles could manifest themselves and influence the values of the observerbles considered in this study. Therefore, the precise measurements of these observables at the HL-LHC and the future $e^+e^-$ colliders provide a valuable opportunity to test the SM and investigate potential new physics beyond the SM.

\appendix

\section*{Appendix: Polarization Vectors and Kinematics}

The general definitions of longitudinal, transverse and normal lepton polarization four-vectors are:
\begin{eqnarray}
    S_L \equiv \left(0,\frac{\mathbf{p_1}}{|\mathbf{p_1}|}\right),\quad S_N \equiv \left(0,\frac{\mathbf{k}\times\mathbf{p}_1}{|\mathbf{k}\times\mathbf{p}_1|}\right),\quad
    S_T\equiv\left(0,\frac{\mathbf{p_1}\times(\mathbf{k}\times\mathbf{p}_1)}{|\mathbf{p_1}\times(\mathbf{k}\times\mathbf{p}_1)|}\right).\notag
\end{eqnarray}

In the lepton rest frame, the above definitions are modified to
\begin{eqnarray}
    S_L = (0,0,0,1),\quad S_N = (0,\sin\phi,-\cos\phi,1),\quad
    S_T=(0,\cos\phi,\sin\phi,1).\notag
\end{eqnarray}

In any frame, moving in the $z$-direction with respect to the lepton, the general definitions are modified using the Lorentz boost $\boldsymbol{\beta}=\frac{\bold{p_1}}{E_1}$,
\begin{eqnarray}
    S'_L = (\frac{\mathbf{p_1}}{m},0,0,\frac{E_1}{m}),\quad S'_N = (0,\sin\phi,-\cos\phi,1),\quad
    S'_T=(0,\cos\phi,\sin\phi,1),\notag
\end{eqnarray}
where $\phi$ is azimuthal angle.
The four momenta of Higgs, lepton and photon in Higgs' rest frame are defined as
\begin{eqnarray}
   p = m_H(1,0,0,0),\quad k = E_\gamma(1,0,0,1),\quad
     p_1 = (E_1,\bold{p_1}\sin\theta\cos\phi,\bold{p_1}\sin\theta\sin\phi,\bold{p_1}\cos\theta).\notag
\end{eqnarray}

The photon polarization vectors are defined as
\begin{eqnarray}
   \epsilon_+&=& \frac{1}{\sqrt{2}}(0,-1,- i,0),   \quad \epsilon_-= \frac{1}{\sqrt{2}}(0,1,- i,0),\quad \epsilon_\pm\cdot\epsilon^{*}_\pm=-1.\notag
\end{eqnarray}
The following scalar products are used in our calculations
\begin{eqnarray}
  p\cdot\epsilon_\pm=k\cdot\epsilon_\pm=0,\quad p_1\cdot\epsilon_\pm= \pm\frac{\bold{p_1}\sin\theta}{\sqrt{2}}e^{\pm i\phi}, \quad
   p_1\cdot\epsilon^{*}_\pm= \pm\frac{\bold{p_1}\sin\theta}{\sqrt{2}}e^{\mp i\phi}.\notag 
\end{eqnarray}
The Levi-Civita symbols used in our calculations are defined as
\begin{eqnarray}
   \epsilon^{k\epsilon_{\pm}p_1p_2}&=&\epsilon^{k\epsilon_{\pm}p_1p},\quad\epsilon^{k\epsilon_{\pm}p_1p_2}= \frac{-i}{\sqrt{2}}m_H \bold{p_1} E_\gamma\sin\theta e^{\pm i\phi},\quad \epsilon^{k\epsilon^{*}_{\pm}p_1p_2}= \frac{i}{\sqrt{2}}m_H \bold{p_1} E_\gamma\sin\theta e^{\mp i\phi},\notag \\ \epsilon^{k\epsilon_\pm \epsilon_\pm ^{*}p}&=& \pm i  E_\gamma m_H, \quad \epsilon^{\epsilon_\pm \epsilon_\pm ^{*}p_1p}= \pm i m_H \bold{p_1} \cos\theta,\quad\epsilon^{k\epsilon_\pm \epsilon_\pm^{*}p_1}= \pm i E_\gamma(E_1-\bold{p_1}\cos\theta).\notag
\end{eqnarray}

\bibliographystyle{apsrev4-1}

\bibliography{main.bib}

\end{document}